\documentclass[a4paper,onecolumn,superscriptaddress,11pt,accepted=2020-05-12]{quantumarticle}
\pdfoutput=1
\usepackage[utf8]{inputenc}
\usepackage[english]{babel}
\usepackage[T1]{fontenc}
\usepackage{amsmath}
\usepackage{hyperref}
\usepackage{enumerate}
\usepackage{doi}

\usepackage[sort&compress,numbers]{natbib}

\usepackage{amssymb}
\usepackage{mathtools}
\usepackage{color}
\usepackage{bbm}
\usepackage{subfig}
\usepackage[]{graphicx}
\usepackage{floatrow}

\makeatletter
\DeclareFontFamily{OMX}{MnSymbolE}{}
\DeclareSymbolFont{MnLargeSymbols}{OMX}{MnSymbolE}{m}{n}
\SetSymbolFont{MnLargeSymbols}{bold}{OMX}{MnSymbolE}{b}{n}
\DeclareFontShape{OMX}{MnSymbolE}{m}{n}{
    <-6>  MnSymbolE5
   <6-7>  MnSymbolE6
   <7-8>  MnSymbolE7
   <8-9>  MnSymbolE8
   <9-10> MnSymbolE9
  <10-12> MnSymbolE10
  <12->   MnSymbolE12
}{}
\DeclareFontShape{OMX}{MnSymbolE}{b}{n}{
    <-6>  MnSymbolE-Bold5
   <6-7>  MnSymbolE-Bold6
   <7-8>  MnSymbolE-Bold7
   <8-9>  MnSymbolE-Bold8
   <9-10> MnSymbolE-Bold9
  <10-12> MnSymbolE-Bold10
  <12->   MnSymbolE-Bold12
}{}

\let\llangle\@undefined
\let\rrangle\@undefined
\DeclareMathDelimiter{\llangle}{\mathopen}%
                     {MnLargeSymbols}{'164}{MnLargeSymbols}{'164}
\DeclareMathDelimiter{\rrangle}{\mathclose}%
                     {MnLargeSymbols}{'171}{MnLargeSymbols}{'171}
\makeatother

\newcommand{\tr}[1]{\text{tr}\left[ #1 \right]}

\newcommand{\qav}[1]{\left<#1\right>}
\newcommand{\ket}[1]{\left\vert#1\right\rangle}
\newcommand{\bra}[1]{\left\langle#1\right\vert}
\newcommand{\ketbra}[2]{\ket{#1}\hspace{-3pt}\bra{#2}}
\newcommand{\com}[1]{\left[ #1 \right]}
\newcommand{\acom}[1]{\left\{ #1 \right\} } 
\newcommand{\bbra}[1]{\left\llangle #1 \right|  } 
\newcommand{\kket}[1]{\left| #1 \right\rrangle } 
\newcommand{\qqav}[1]{\left\llangle #1 \right\rrangle}

\DeclareMathAlphabet\mathbfcal{OMS}{cmsy}{b}{n}
\DeclareMathAlphabet{\mathcal}{OMS}{cmsy}{m}{n}

\begin{document}

\title{Time crystallinity in open quantum systems}
\date{\today}
\author{Andreu Riera-Campeny}
\affiliation{F\'{\i}sica Te\`{o}rica: Informaci\'{o} i Fen\`{o}mens Qu\`{a}ntics. Departament de F\'{\i}sica, Universitat Aut\`{o}noma de Barcelona, 08193 Bellaterra, Spain}
\email{andreu.riera.campeny@uab.cat}
\author{Mariona Moreno-Cardoner}
\affiliation{F\'{\i}sica Te\`{o}rica: Informaci\'{o} i Fen\`{o}mens Qu\`{a}ntics. Departament de F\'{\i}sica, Universitat Aut\`{o}noma de Barcelona, 08193 Bellaterra, Spain}
\author{Anna Sanpera}
\affiliation{F\'{\i}sica Te\`{o}rica: Informaci\'{o} i Fen\`{o}mens Qu\`{a}ntics. Departament de F\'{\i}sica, Universitat Aut\`{o}noma de Barcelona, 08193 Bellaterra, Spain}
\affiliation{ICREA, Passeig Llu\' is Companys 23, 08001 Barcelona, Spain.}

\maketitle

\begin{abstract}
Time crystals are genuinely non-equilibrium quantum phases of matter that break time-translational symmetry. While in non-equilibrium closed systems time crystals have been experimentally realized, it remains an open question whether or not such a phase survives when systems are coupled to an environment. Although dissipation caused by the coupling to a bath may stabilize time crystals in some regimes, the introduction of incoherent noise may also destroy the time crystalline order. Therefore, the mechanisms that stabilize a time crystal in open and closed systems are not necessarily the same. Here, we propose a way to identify an open system time crystal based on a single object: the Floquet propagator. Armed with such a description we show time-crystalline behavior in an explicitly short-range interacting open system and demonstrate the crucial role of the nature of the decay processes.
\end{abstract}

\section{Introduction and outline}\label{sec:intro}

Statistical mechanics has been extremely successful in describing the behavior of systems at equilibrium and, occasionally, even the relaxation towards it. During the last decades, countless efforts have been devoted to genuinely non-equilibrium systems. In particular, a lot of attention has been drawn to non-equilibrium \textit{Floquet} systems, i.e., systems undergoing time-periodic dynamics. Those systems have found numerous applications that go from thermal machines and transport \cite{Kosloff2013,Alicki2015,Restrepo2018,Niedenzu2018,Riera2019}, to \textit{Floquet engineering} \cite{Bukov2015,Cayssol2013,DelCampo2013}, as well as the discovery of non-equilibrium phases of matter \cite{Keyserlingk2016, Khemani2017, Else2016, Else2017, Zhang2017, Choi2017_2,Choi2017_1,Yao2017,Gong2018,Iemini2018}. This work focuses on the latter, the so-called time crystals. 

A system has \textit{discrete time-translational symmetry} if the generator of the evolution, at any time $t$, is invariant under the transformation $t \mapsto t+T$, where $T$ is the period. The pioneering ideas of time-crystals proposed in \cite{Wilczek2012} and polished by subsequent discussions in \cite{Bruno2013}, led to the concept of \textit{discrete time crystals} (DTCs), first put forward in \cite{Else2016,Choi2017_1} and then experimentally realized in  \cite{Zhang2017,Choi2017_2}. A discrete time crystal is a {\color{black} many-body} system that breaks discrete time-translational symmetry showing robust subharmonic response (to be precised below). Closed quantum systems might display oscillations in a wide variety of scenarios; e.g., from Rabi oscillations of quantum optical systems to Bloch oscillations in lattices. Hence, criteria on how to identify a time crystal are essential to understand this non-equilibrium phase of nature. In \textit{closed} systems, a discrete time crystal phase is characterized by an observable $\text{O}$ acting as an \textit{order parameter} whose expectation value $O(t) = \tr{\text{O} \rho(t)}$ must fullfill the three following conditions: \cite{Khemani2017, Russomanno2017, Huang2018}: 
\begin{enumerate}[(I)]
\item \textbf{Time-translation symmetry breaking:} the order parameter is less symmetric than the Hamiltonian, i.e., $O(t+T) \neq O(t)$ when $\text{H}(t+T) = \text{H}(t)$. For the DTC we recover $O(t)$ after an integer number of periods $N>1$, $O(t+NT) = O(t)$.
\item \textbf{Rigidity of the oscillations:} $O(t)$ shows a fixed oscillation period $NT$ without fine-tuned Hamiltonian parameters. Equivalently, the oscillations should \textit{lock-off} at frequency $2\pi/(NT)$. 
\item \textbf{Persistence to infinite time:} the non-trivial oscillation with period $T$ must persist for infinitely long time in the thermodynamic limit. 
\end{enumerate}
Nevertheless, it is not clear whether this phase of matter survives the action of an environment or, even more interesting, if the environment can help to stabilize it \cite{Else2017}. For instance, the discrete time crystal appearing in a disordered one dimensional Ising spin chain (the so-called $\pi$SG) cannot survive the coupling to an environment \cite{Lazarides2017}. As a result, efforts have been redirected {\color{black} mostly} towards the study of open \textit{mean-field} models with collective interactions \cite{Iemini2018, Gambetta2019, Gong2018} or short-ranged perturbations from them \cite{Zhu2019} coupled to an environment. In such models, some signatures of time crystallinity have been theoretically predicted, but there is a still a controversy on whether the mean-field description used on those models could sweep under the carpet part of the relevant physics which would destroy the time crystalline order. Part of the controversy is related to the fact that there is no well-posed description of what an \textit{open system} time-crystal should be. 

In this work we propose plausible criteria to define and characterize discrete time crystals in open systems governed by a Lindblad master equation. Our study is based on the analysis of the so-called Floquet propagator and the properties the associated Liouvillian must have in order to support time crystallinity. Further, we investigate the stability mechanisms and analyze how time crystals can be implemented in open systems. Also, there has been some discussion around the possibility that only \textit{mean-field} models can exhibit time crystalline order in open quantum systems. Only very recently, the possibility of having time crystals in an open Ising model have been discussed \cite{Lazarides2019}. Our analysis shows that collective interactions are not crucial features to observe time-crystalline behavior. Nonetheless, our findings show that \textit{collective} decay processes are relevant in order to have subharmonic oscillations that are more robust to errors. 

The outline of the article is as follows: In Sec.~\ref{sec:preliminaries}, we present in detail the most relevant tools and concepts used throughout. In Sec.~\ref{sec:time-crystals}, we introduce the definition of time crystals in open systems and derive some important properties to further characterize them. In section Sec.~\ref{sec:intuition}, a collection of low dimensional open system examples is introduced in order to build up some intuition on how to reach stability of time crystals.  Sec.~\ref{sec:openXY} and Sec.~\ref{sec:results} form the main body of this work, there we present a short range many-body model (XY model) and evaluate its properties and validity as an open system time crystal using the definitions proposed in Sec.~\ref{sec:time-crystals}.  To this aim we derive its corresponding master equation and solve it numerically. Finally, we present our conclusions. 
{\color{black} An expert reader which is not interested in details may go directly to Sec.~\ref{sec:time-crystals} and Sec.~\ref{sec:results}, where the definition of the open system time-crystals and the results for our case of study are respectively given. }


\section{Open quantum system dynamics: concepts and tools}\label{sec:preliminaries}

In this section, we introduce the tools of --Markovian-- open quantum systems and fix the notation used throughout the article. In what it follows, $\mathcal{H}$ denotes a Hilbert space of dimension $d_\mathcal{H}$. {\color{black} States are positive and trace-one operators that we denote by Greek symbols, for instance, $\rho\in \text{S}(\mathcal{H})$ being $\text{S}(\mathcal{H})$ the set of states. The set of operators in this Hilbert space is $\text{Op}(\mathcal{H})$, and its elements are denoted by regular text characters $\text{A}\in\text{Op}(\mathcal{H})$. Likewise, the set of \textit{superoperators}, i.e., linear maps between operators, are denoted with calligraphic symbols $\mathcal{A}\in\text{SOp}(\mathcal{H})$.}

\subsection{Open systems dynamics}\label{subsec:open_systems}

Consider a closed time-independent physical system whose dynamics is generated by a Hamiltonian $\text{H} \in \text{Op}(\mathcal{H})$. Then, the Schrödinger equation:
\begin{align}
\partial_t \ket{\Phi} = -i \text{H}\ket{\Phi} \rightarrow \ket{\Phi(t)} = e^{-i \text{H}t} \ket{\Phi(0)}, \label{eq:unit_evo}
\end{align}
and the state $\ket{\Phi}$ undergoes unitary dynamics. Often, one is interested only on the dynamics of a reduced set of the degrees of freedom commonly referred to as \textit{system} (S). Complementary to those, there are \textit{bath} (B) or \textit{environment} degrees of freedom. A partition $\text{H} =  \text{H}_S +\text{H}_{SB} +\text{H}_{B}$ is always possible, where the subscript indicates the degrees of freedom of the system ($S$),  the bath ($B$) or both at the same time ($SB$). The Markovian evolution of a reduced system $S$ is generated by the so-called Lindblad master equation: 
\begin{align}
\partial_t \rho = \mathcal{L}(\rho) = - i \com{\text{H}_{S},\rho} + \sum_\alpha \left( \text{L}^\alpha \rho \text{L}^{\alpha \dagger} -\frac{1}{2}\acom{\text{L}^{\alpha \dagger} \text{L}^\alpha,\rho } \right). \label{eq:lindblad}
\end{align}
where $\mathcal{L} \in \text{SOp}(\mathcal{H})$ is the \textit{Liouvillian} superoperator and $\text{L}^\alpha$ are the jump operators. We assume the Liouvillian to be time-independent unless otherwise stated. When the state of the system is initially uncorrelated from that of the environment, Eq.~\eqref{eq:lindblad} can be derived from Eq.~\eqref{eq:unit_evo} under three key approximations \cite{Breuer2002,Alicki2007}: the coupling between the system and the environment is weak (weak-coupling approximation), and the environment equilibrates fast (Markov approximation). In addition, the fast rotating terms are usually disregarded (secular approximation). Note that Eq.~\eqref{eq:lindblad} is a hermiticity preserving equation, i.e., $\mathcal{L}(\rho^\dagger) = (\mathcal{L}(\rho))^\dagger$. Such evolution generates a family of completely-positive and trace-preserving (CPTP) maps of the form $\mathcal{E}(t) = \exp(\mathcal{L}t)$. The superoperator $\mathcal{E}(t)$ is known as the evolution map and satisfies $\mathcal{E}(t+t') = \mathcal{E}(t)\circ \mathcal{E}(t')$ and $\mathcal{E}(0) = \mathcal{I}$, where $\mathcal{I}$ is the identity map. 

Since linear operators form a vector space, {\color{black} it is possible to represent them as vectors of a larger Hilbert space using a procedure known as vectorization. Given a basis $\{\ket{i}\}$ of $\mathcal{H}$, the vectorization consists in, essentially, the replacement $\text{A}=\sum_{ij}\text{A}_{ij} \ket{i}\bra{j}\in \text{Op}(\mathcal{H}) \mapsto  \kket{\text{A}} =\sum_{ij}\text{A}_{ij} \ket{i}\otimes \ket{j}^*\in \mathcal{H}\otimes\mathcal{H}$. Accordingly, a product of the form $\text{A} \rho \text{B}$ should be replaced by $\text{A} \otimes \text{B}^\text{T} \kket{\rho}$. Hence, the same transformation holds for linear superoperators, like the Liouvillian $\mathcal{L}$, being now regular operators on $\mathcal{H}\otimes\mathcal{H}$, i.e. $\mathcal{A}(\rho) \mapsto \mathcal{A}\kket{\rho}$.} The inner product in the extended Hilbert space is given by the Hilbert-Schmidt product defined by $\qqav{\text{A}|\text{B}} = \tr{\text{A}^\dagger \text{B}}$. It automatically introduces the notion of \textit{adjoint} superoperator which, for all $\text{A},\text{B}\in \text{Op}(\mathcal{H})$ fulfills
\begin{align}
\qqav{ \text{A}| \mathcal{A}(\text{B})} = \tr{\text{A}^\dagger \mathcal{A} (\text{B})} = \tr{ ( \mathcal{A}^\ddagger(\text{A}))^\dagger \text{B}} = \qqav{\mathcal{A}^\ddagger (\text{A})| \text{B}},
\end{align} 
with the property $\mathcal{A}(\cdot) = \text{B} \cdot \text{C} \Rightarrow \mathcal{A}^\ddagger (\cdot) =  \text{B}^\dagger \cdot \text{C}^\dagger$. In particular, the adjoint Lindblad equation yields
\begin{align}
\mathcal{L}^\ddagger(\text{A}) =  i \com{\text{H}_\text{S},\text{A}} + \sum_\alpha \left( \text{L}^{\alpha\dagger}\com{\text{A}, \text{L}^{\alpha}} + \com{\text{L}^{\alpha\dagger},\text{A} }\text{L}^{\alpha} \right). \label{eq:adjoint_lindblad}
\end{align}
{\color{black} In essence, the vectorization procedure provides a matrix representation of the Liouvillian $\mathcal{L}$. Therefore, some features of the evolution can be learned from its spectral decomposition, which is the objective of the following subsection. }

\subsection{Spectral properties of the Liouvillian}

{\color{black} A way to grasp the properties of the dynamics of an open system is by analyzing the spectrum of the Liouvillian matrix $\mathcal{L}$. We focus here in the case of a diagonalizable Liouvillian and  refer the interested reader to App.~\ref{app:generator_math} and references \cite{Lindblad1976, Puri2001, Breuer2002, Baumgartner2008_1} for an extended discussion of the non-diagonalizable case.} The set of $d_\mathcal{H}^2$ eigenvalues $\{\lambda_\mu\}$ are found as the roots of the characteristic polynomial $\text{P}_\mathcal{L}(\lambda) = \text{det}(\mathcal{L}-\lambda\mathcal{I})$. The corresponding ordinary \textit{left} and \textit{right} eigenvectors are defined as non-trivial solutions of the equations
\begin{align}
\mathcal{L}^\ddagger \kket{\text{l}_\mu}  = \lambda_\mu^* \kket{\text{l}_\mu},\nonumber \\
\mathcal{L} \kket{\text{r}_\mu} = \lambda_\mu \kket{\text{r}_\mu}\label{eq:right_evec} .
\end{align}
{\color{black} The spectral decomposition above has the following properties :}

\begin{figure}
  \sidesubfloat[]{\includegraphics[width=0.44\textwidth]{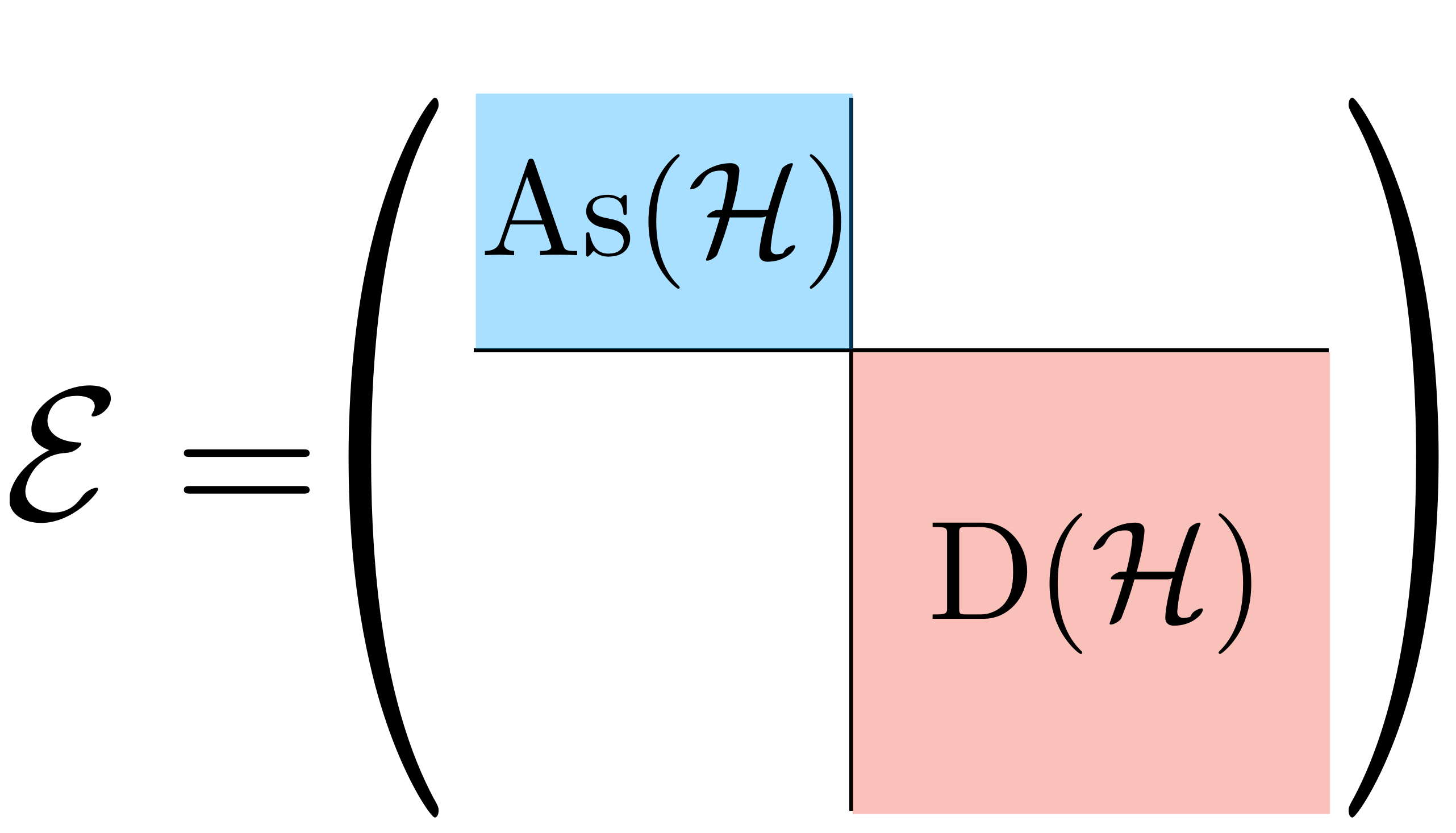}\label{fig:fig1a}}\quad
  \sidesubfloat[]{\includegraphics[width=0.44\textwidth]{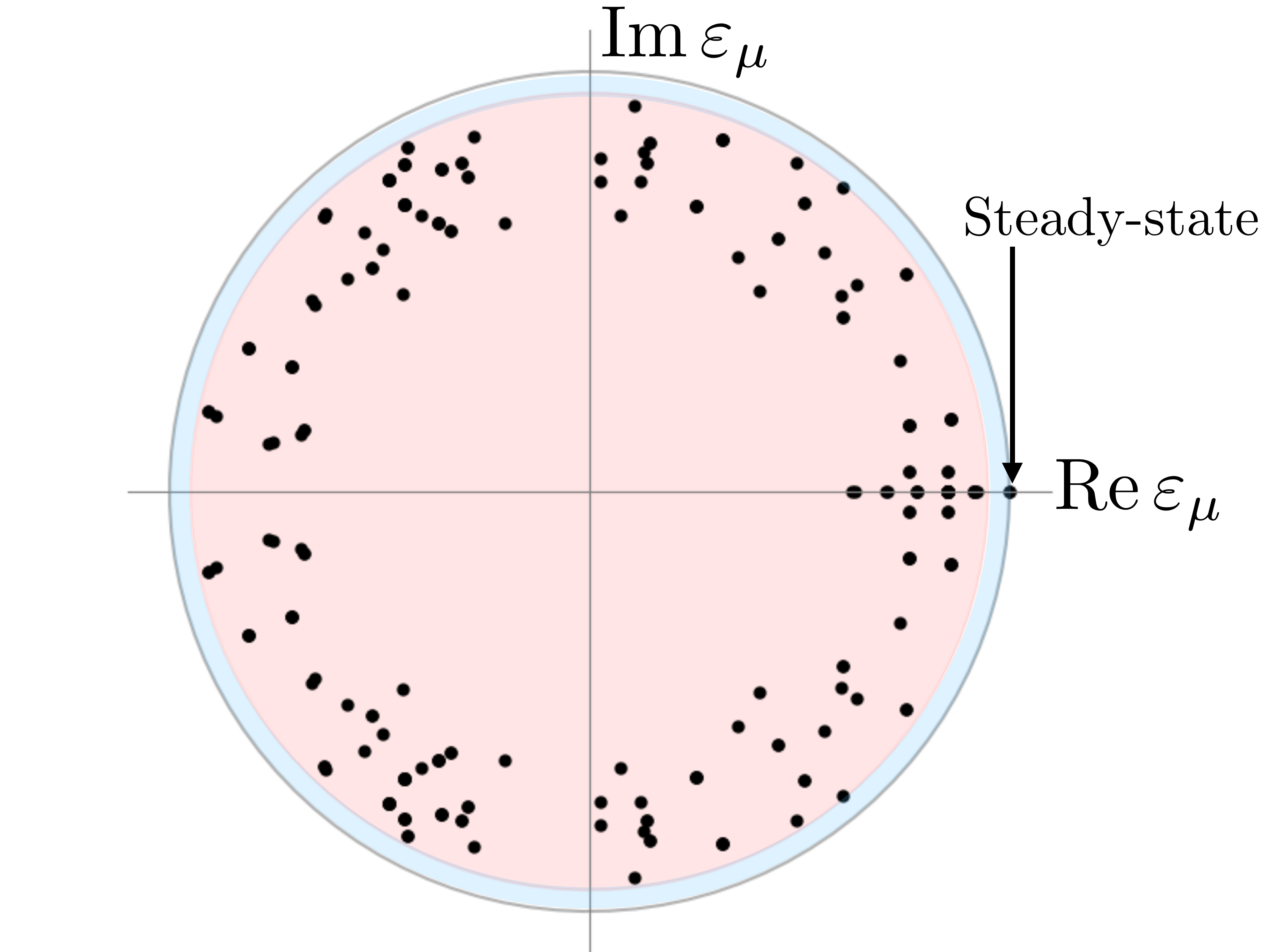}\label{fig:fig1b}}
  \caption{(a) Schematic decomposition of $\mathcal{H}\otimes\mathcal{H}$ in terms of the asymptotic (blue) and decay (red) subspaces. (b) Typical spectrum of a CPTP map where the peripheral spectrum is colored in blue and the decaying subspace in red. \label{fig:fig1}}
\end{figure}

\begin{enumerate}[(i)]
\item The eigenvalues of $\mathcal{L}$ are either real or come by conjugate pairs. Also, the positivity of the evolution requires the eigenvalues to have negative real part $\text{Re}\lambda_\mu \leq 0$.
\item Ordinary eigenvectors {\color{black} corresponding to} different eigenvalues are linearly independent. 
\item The ordinary eigenvectors of $\mathcal{L}$ and $\mathcal{L}^\ddagger$ can be chosen bi-orthogonal, i.e. $\qqav{\text{l}_\mu|\text{r}_\nu} =\delta_{\mu\nu}$. More compactly, for diagonalizable $\mathcal{L}$, we have $\mathcal{W}_\text{l}^\ddagger \mathcal{W}_\text{r} = \mathcal{I}$, where the columns of $\mathcal{W}_r$ are the right eigenvectors $\kket{\text{r}_\mu}$ (and similarly for $\mathcal{W}_l$). 
\item  The evolution map has the form $\mathcal{E}(t) = \exp(\mathcal
{L}t)$ and, therefore, $\mathcal{E}(t)$ and $\mathcal{L}$ share the same left and right eigenvectors. 

\item For any {\color{black} time-independent} $\mathcal{L}$ there is always one eigenvalue $\lambda_0=\lambda^*_0=0$, with left eigenvector $\bbra{\text{1}}$. The corresponding right eigenvector $\kket{\text{r}_0}$ fulfills $\mathcal{E}(t)\kket{\text{r}_0} = \kket{\text{r}_0}$ and is often referred to as the steady-state. {\color{black} However, note that there may be other \textit{non-decaying} eigenvectors $\kket{\text{r}_\mu}$ such that $\text{Re}\lambda_\mu = 0$}. 
\end{enumerate}
{\color{black} The possibility of having multiple non-decaying states, as stated in property (v), is often disregarded. This possibility implies that the asymptotic state of a system, i.e. $\rho(t\to\infty)$, is not unique. To capture this asymptotic behavior}, we introduce the notion of asymptotic subspace:
\begin{align}
\text{As}(\mathcal{H}) = \text{span}\acom{\text{r}_\mu : \text{Re}{\lambda_\mu} =0},
\end{align}
{\color{black}which will become of crucial importance to this work. From now on, we denote by $\Psi_\mu$ a general element of $\text{As}(\mathcal{H})$. Some coments are in order: First, we remark that $\Psi_\mu$ are general elements of $\text{Op}(\mathcal{H})$ and not always proper quantum states. Also, the elements $\Psi_\mu$ are non-decaying rather than steady, since a non-zero imaginary part $\text{Im}\lambda_\mu\neq 0$ causes them to gain a time-dependent phase.} It can be proven, see for instance \cite{Albert2014}, that the asymptotic subspace can be always diagonalized, otherwise the dynamics would explode as $t\to\infty$. {\color{black} We refer to the subspace orthogonal to $\text{As}(\mathcal{H})$ as the decay space $\text{D}(\mathcal{H})$ (see Fig.\ref{fig:fig1a})}. Finally, we also define the \textit{dissipative gap} $\Delta$ as:
\begin{align}
\Delta \coloneqq \min_{\mu} |\text{Re}\lambda_\mu| \quad \text{such that } \quad \text{Re}\lambda_\mu\neq 0,
\end{align}
{\color{black}which is the relevant quantity }that fixes the time-scale of convergence towards the {\color{black} asymptotic} state of the system. 

    \floatsetup[figure]{style=plain,subcapbesideposition=top}

\subsection{Conserved quantities of the evolution}

In closed systems, any symmetry of the Hamiltonian is a conserved quantity of the evolution. In dissipative systems, this is not always the case, and the relation between symmetries and conserved quantities is, in general, more complex. A correspondance between asymptotic states and conserved quantities was presented in \cite{Albert2016}. Essentially, given $\Psi_\mu \in \text{As}(\mathcal{H})$ with purely imaginary eigenvalues $\lambda_\mu = i \phi_\mu$, there is a corresponding conserved quantity $\text{j}_\mu\in \text{Op}(\mathcal{H})$ such that, for any initial state $\rho(0)$, the asymptotic state is given
\begin{align}
\rho(t) = \sum_{\mu} e^{i \phi_{\mu} t}  j_\mu \Psi_{\mu} + \mathcal{O}(e^{-\Delta t}),\label{eq:rho_asymptotic}
\end{align}
where $j_\mu = \text{tr}[\text{j}_{\mu}^\dagger \rho(0)]$ is the imprint of the initial state on the conserved quantities. For instance, when there is only one asymptotic state, $\text{j}_0 = \text{1}$ is the only conserved quantity and its expectation value $j_0 = \tr{\rho(0)} = 1$ is preserved throughout the evolution. In \cite{Albert2016}, the authors found an analytic expression for the conserved quantities $\text{j}_\mu$ in terms of, essentially, the Liouvillian $\mathcal{L}$ and the asymptotic states $\Psi_\mu$. This relation is used throughout Sec.~\ref{sec:intuition} to compute the conserved quantities in some exemplary models.

\section{Discrete time crystals beyond closed systems}\label{sec:time-crystals}

{\color{black} 
In the former section, we have discussed the evolution of systems under a time-independent Liouvillian $\mathcal{L}$. We are ultimately interested in time crystals and, therefore, the discussion should be extended to time-periodic evolutions $\mathcal{L}(t+T) = \mathcal{L}(t)$. For time-dependent systems, the evolution map is given by the well-known Dyson solution:
\begin{align}
\mathcal{E}(t) = \mathcal{T}\exp\left( \int_0^t ds \mathcal{L}(s) \right),
\end{align}
where $\mathcal{T}$ is the time-ordering operator. Note that, in this situation, the spectral properties of the instantaneous $\mathcal{L}(t)$ are no longer relevant. For time-periodic systems, the evolution map for a period $T$
\begin{align}
\mathcal{E}_\text{F} = \mathcal{T}\exp\left( \int_0^T ds \mathcal{L}(s)\right), \label{floquet_propagator}
\end{align}
is known as the Floquet propagator \cite{Gong2018}, and encodes the information about the stroboscopic evolution of the system. Namely, for times  $t_n = nT$ with $n\in\mathbb{N}$, the evolution map is $\mathcal{E}(t_n) =  \mathcal{E}_\text{F}^n = \mathcal{E}_\text{F}\circ\cdots \circ \mathcal{E}_\text{F}$. The map $\mathcal{E}_\text{F}$ is positive and, therefore, its eigenvalues lie within the unit disk. Moreover, the trace preserving condition guarantees at least one eigenvalue equal to one. In this picture, the asymptotic subspace corresponds to the span of eigenvectors whose eigenvalues lie on the radius one circle (see Fig.\ref{fig:fig1b}). For this reason, the set of eigenvalues associated to the asymptotic subspace is termed \textit{peripheral spectrum}. The following question arises: can we use the spectrum of $\mathcal{E}_\text{F}$ to characterize time crystals in open systems? The answer to this question is positive and we formalize it below.}

As discussed in the introduction, time crystals in closed systems are identified by exhibiting: (I) discrete \textbf{time-translational symmetry breaking}, (II) \textbf{ rigidity} on the subharmonic response of the order parameter and (III) the infinite \textbf{persistence} of the subharmonic response in the thermodynamic limit.
Based in these conditions, we propose to characterize an \textit{open system time crystal} using a single object, the Floquet propagator $\mathcal{E}_\text{F}$. {\color{black} We denote the eigenvalues of $\mathcal{E}_\text{F}$ as $\{\varepsilon_\mu\}$ and, again, the left and right eigenvectors are 
\begin{align}
\mathcal{E}_\text{F}^\ddagger \kket{\text{l}_\mu}  = \varepsilon_\mu^* \kket{\text{l}_\mu},\nonumber \\
\mathcal{E}_\text{F} \kket{\text{r}_\mu} = \varepsilon_\mu \kket{\text{r}_\mu}.
\end{align}
The associated Floquet asymptotic subspace corresponds now to $\text{As}(\mathcal{H})_\text{F} = \text{span}\left\{ \text{r}_\mu : |\varepsilon_\mu| = 1\right\}$ and again we denote its elements by $\Psi_\mu$. With these definitions settled, a many-body system can be identified as an open time crystal if its Floquet propagator fulfills:}
\begin{enumerate}[(I)]
\item \textbf{Time-translation symmetry breaking:} there exists at least one {\color{black} eigenvalue $\varepsilon_\star \in \{\varepsilon_\mu\}$ such that $\varepsilon_\star \neq 1$ but $\varepsilon_\star^N = 1$ for some integer $N$}. 
\item \textbf{Rigidity of the oscillations:} {\color{black} In the thermodynamic limit, the eigenvalue $\varepsilon_\star$ is linearly robust. Namely, given a deformation of the evolution $\mathcal{E}_\text{F} \mapsto \mathcal{E}_\text{F} + \eta \mathcal{V}$ the \textit{susceptibility} $\chi^{(1)} = \left|\left(\partial\varepsilon_{\mu^\star}/\partial \eta \right)_{\eta=0}\right| = 0$.}
\item  \textbf{Persistence of the oscillations to infinite time:} The time-scale of decay of the subharmonic oscillations is fixed by the \textit{dissipative Floquet gap} $\Delta_\text{F} = (-1/T) \log |\varepsilon_\star|$, which is zero if $\varepsilon_\star$ is in the peripheral spectrum.
\end{enumerate}
A system fulfilling (I)--(III) displays, in the thermodynamic limit, infinitely-lived and robust subharmonic response and, therefore, it is in a \textit{time-crystal phase}. In the opposite case, the system typically {\color{black}equilibrates} reaching the \textit{thermal phase}.

{\color{black} 
We remark that our definition of the time-crystal phase, specially condition (II), is done in a \textit{weak} sense. Namely, for equilibrium phases of matter, one requires absolute (i.e., to all orders) stability against perturbations in the thermodynamic limit. In contrast, condition (II) is a necessary condition of absolute stability but, obviously, not sufficient. However, from a practical perspective, the difference between the linear and the absolute stability conditions will only be apparent for presumably very long times. Therefore, we consider here only the linear stability and leave the tightening of condition (II) for future research.}\\
{\color{black} Let us illustrate the behavior of a system in a time-crystal phase, with subharmonic oscillations of periodicity $2T$. In this case, the Floquet propagator has two eigenvalues $\varepsilon_0=1$ and $\varepsilon_\star = e^{i\pi}$. The state of the system after a large number of oscillations} $n$ is well approximated, using Eq.~\eqref{eq:rho_asymptotic}, by 
\begin{align}
{\color{black} \lim_{n\to\infty} \rho(nT)} = {\color{black} \lim_{n\to\infty} } \mathcal{E}_F^{n}( \rho(0)) \approx \Psi_0 + (-)^n j_\star \Psi_\star, 
\end{align}
where we used $j_0 = \text{tr}[\rho(0)] =1$. As a result, the expectation value of the order parameter after $n$ periods {\color{black} is
\begin{align}
O(nT) &= \text{tr}[\text{O} \mathcal{E}_\text{F}^n (\rho(0))] \approx \text{tr}[\text{O} \Psi_0]+(-)^n j_\star \text{tr}[\text{O} \Psi_\star].
\end{align}
Therefore, one observes long-lived subharmonic oscillations if: 
\begin{enumerate}[(a)]
\item the choice of the initial state $\rho(0)$ is such that $\text{tr}[\text{j}_\star^\dagger \rho(0)] \neq 0$, i.e., the initial state has some overlap with the conserved quantity $\text{j}_\star$, and
\item the choice of the order parameter $\text{O}$ is such that $\text{tr}[\text{O}\Psi_\star] \neq 0 $, i.e., the order parameter is sensible to the subharmonic oscillations. 
\end{enumerate}}

\subsection{Open systems and the kicked protocol}

{\color{black} Before proceeding to further study the implications of our definition of the time-crystal phase, it is timely to motivate our choice of the driving protocol $\mathcal{L}(t)$. In general, an analytically tractable expression of $\mathcal{E}_\text{F}$ is impossible to obtain. In particular, starting from a microscopic system-bath model, the computation of $\mathcal{E}_\text{F}$ demands, as a first step, finding the time-ordered exponential of a time-periodic Hamiltonian. This is already a difficult problem on its own (see for instance \cite{Dalibard2014,Eckardt2015}). In this work, we focus in what we call \textit{kicked protocols}, for which the computation of $\mathcal{E}_\text{F}$ is analytically doable. We start by considering a total Hamiltonian: 
\begin{align}
&\text{H}(t) = \text{H}_S + \text{H}'_S(t) + \text{H}_{SB}+ \text{H}_B = \text{H} + \text{H}_S'(t)\\
&\text{H}'_S(t) =  g\sum_n \delta(t-nT)\text{H}_K,
\end{align}
where $\text{H}$ is explicitly time-independent and $n\in\mathbb{N}$. Then, for $\sigma(0) = \rho(0)\otimes\rho_B$ the evolution over one period of the system is
\begin{align}
\rho(T) &= \text{U}_\text{K} \text{tr}_\text{B} \left[ e^{-i\text{H} T} \sigma(0) e^{i \text{H} T} \right] \text{U}^\dagger_\text{K} \Rightarrow \rho(T) = \mathcal{E}_\text{F}(\rho(0)) = \text{U}_\text{K} \left( e^{\mathcal{L}T} \rho(0) \right) \text{U}^\dagger_\text{K},\label{eq:kicked_open_system}
\end{align}}
where $\mathcal{L}$ is the Liouvillian of the static dissipative evolution, and $\text{U}_\text{K} = \exp(-i g\text{H}_\text{K})$ is the unitary kick operator. {\color{black} Therefore, for a kicked open system, one can use the dissipation model of the time-independent problem, with the dynamics being interrupted periodically by the unitary kick.} Note that, the Born approximation guarantees the state of the bath to remain unchanged after one period of the evolution. Therefore, we can iterate this process to obtain the state of the system at stroboscopic times $t_n$. The kicked protocol is used, for instance, in the models studied in \cite{Gong2018, Gambetta2019, Zhu2019}.

\subsection{Useful observations for open system time crystals}\label{subsec:observations}

{\color{black} In view of the relevance of the Floquet propagator $\mathcal{E}_\text{F}$, we here analyze some of its properties. We focus on the kicked protocol described in the previous subsection. Namely, a two step evolution consisting on a dissipative dynamics $\mathcal{L}$ during a time $T$, followed by a unitary kick $\text{U}_\text{K}$. The proofs of the following observations can be found in App.~\ref{app:proofs}.}\\

\textbf{Observation 1 (kicked system propagator)}\\
\textit{The Floquet propagator $ \mathcal{E}_\textnormal{F}$ of a system under a kicked protocol takes the simple form:}
\begin{align}
\mathcal{E}_\text{F}(\cdot) = \text{U}_\text{K} \left( e^{\mathcal{L} T} (\cdot)\right)\text{U}_\text{K}^\dagger  \mapsto \mathcal{E}_\text{F} = \text{U}_\text{K} \otimes \text{U}_\text{K}^* \exp\left(\mathcal{L} T\right),\label{eq:kicked_propagator}
\end{align}
\textit{and has always an eigenvalue $\varepsilon_0 = 1$}.\\

{\color{black} Despite the similarity of Eq.~\eqref{eq:kicked_propagator} with a unitary transformation, it is important to note that the spectrum of $\mathcal{L}$ and $\mathcal{E}_\text{F}$ are, in general, not correlated. In observation 2 and observation 3 we detail the relation between the spectra of unitarily connected CPTP maps.}\\

\textbf{Observation 2 (kicked spectrum)}\\
\textit{The spectrum of a  CPTP map $\mathcal{E}$ is generically changed under a kick $\mathcal{U}_\textnormal{K} =\textnormal{U}_\textnormal{K} \otimes \textnormal{U}_\textnormal{K}^*$ corresponding to ${\color{black}\mathcal{E}'} = \mathcal{U}_\textnormal{K} \mathcal{E}$. {\color{black} This is, for instance, the case of $\mathcal{E}_\textnormal{F}$.} }\\

\textbf{Observation 3 (unitarily transformed spectrum)}\\
\textit{The spectrum of a CPTP map $\mathcal{E}$ is unchanged under the unitary transformation $\mathcal{U}_\textnormal{K} =\textnormal{U}_\textnormal{K} \otimes \textnormal{U}_\textnormal{K}^*$ corresponding to ${\color{black}\mathcal{E}''} =  \mathcal{U}_\textnormal{K} \mathcal{E} \mathcal{U}^\ddagger_\textnormal{K}$.}\\

{\color{black} In our time crystal definition, we already introduced that having more than one state in $\text{As}(\mathcal{H})_\text{F}$ is a necessary condition to observe time-translation symmetry breaking. In observation 4, we detail the relation between time crystals and multistability. }\\

\textbf{Observation 4 (multistability and time crystals)}\\
\textit{A CPTP Floquet map $\mathcal{E}_\textnormal{F}$ supports time crystalline behavior only if $\textnormal{dim As}(\mathcal{H})_\text{F} > 1$. The converse is not true, i.e., there are multistable systems without subharmonic oscillations.} \\

{\color{black} Therefore, the structure of $\text{As}(\mathcal{H})_\text{F}$ is crucial in order to observe subharmonic oscillations. Then, knowing that the spectrum of $\mathcal{E}_\text{F}$ differs from that of $\mathcal{L}$, how do their respective asymptotic subspaces relate?}\\

\textbf{Observation 5 (asymptotic space of a kicked evolution)}\\
\textit{Consider $\{\ket{\psi_k}\}$ and $\{\ket{\phi_k}\}$ two basis of $\mathcal{H}$ and a CPTP map $\mathcal{E}$. {\color{black} We denote $S_u$ the subset of basis elements of $\mathcal{H}\otimes\mathcal{H}$ for which the map $\mathcal{E}$ acts unitarily, namely the set $ S_u= \{\ketbra{\psi_k}{\psi_k} : \, \mathcal{E}(\ketbra{\psi_k}{\psi_{k'}}) = \ketbra{\phi_k}{\phi_{k'}}\}$. Then, it exists a unitary kick $\textnormal{U}_\textnormal{K}$ such that $\mathcal{E}_\textnormal{F} = \textnormal{U}_\textnormal{K}\otimes \textnormal{U}_\textnormal{K}^* \mathcal{E}$ has at least $|S_u|$  elements in its asymptotic space.}}\\

{\color{black}We remark that,} in general, observation 5 implies that the dimension of the asymptotic space of a CPTP map $\mathcal{E}$ can increase, decrease or stay equal after a unitary kick. \\

{\color{black} Finally, we present a protocol that exhibits subharmonic response for a CPTP map.} \\

\textbf{Observation 6 (a protocol for sub-harmonic response)}\\
\textit{A general kicked protocol on a CPTP map $\mathcal{E}$ with a unitary $\textnormal{U}_\textnormal{K}$ that gives rise to sub-harmonic response can be obtained by demanding:}
\begin{enumerate}[(i)]
\item the map $\mathcal{E}$ exhibits static multistability, i.e., the peripheral spectrum contains only eigenvalues $\varepsilon_\mu = 1$.
\item the unitary kick $\mathcal{U}_\text{K} = \text{U}_\textnormal{K}\otimes \text{U}_\textnormal{K}^*$ acts independently on the asymptotic and decay spaces, i.e., $\mathcal{U}_\text{K}  = \mathcal{U}_\textnormal{As} \oplus \mathcal{U}_\textnormal{D}$.
\item the unitary kick $\mathcal{U}_\text{As}$ has eigenvalues $u_\alpha = \exp( i  n_\alpha 2\pi/N)$ with $n_\alpha \in \mathbb{Z}$, and at least one eigenvalue is different from one.
\end{enumerate}

{\color{black} Even though the conditions in observation 6 are restrictive, we shall see that they are met for the many-body system studied in Sec.~\ref{sec:openXY}}.

\section{Exemplary: Few-body periodically driven open systems } \label{sec:intuition}

The structure of the asymptotic subspace $\text{As}(\mathcal{H})$ is crucial to identify when a physical system can support subharmonic response. {\color{black} To gather some intuition, we present a collection of one and two qubit models with different structures of $\text{As}(\mathcal{H})$. For each model, we are interested in:
\begin{enumerate}[(i)]
\item finding the asymptotic states $\{\Psi_\mu\}$ and the corresponding conserved quantities $\{\text{j}_\mu\}$,
\item studying the subharmonic response of a kicked protocol,
\item studying the rigidity of the oscillations by perturbing the driving protocol.
\end{enumerate}}
{\color{black} The rigidity must be defined with respect to a given perturbation $\mathcal{V}$. In this work, we pay special attention to rotation errors. Consider a rotation $\text{U}(\theta) = \exp(-i \theta \text{G})$, where $\text{G}$ is the generator of the rotation and $\theta$ the rotation angle. A rotated state is defined as $\rho_\theta=  \text{U}(\theta) \rho_0 \text{U}(\theta)^\dagger$. If the unitary rotation has been implemented imperfectly, namely $\theta \mapsto \theta +\eta$, the perturbation $\mathcal{V}$ is found as
\begin{align}
&\rho_{\theta+\eta}= \text{U}(\theta) \rho_0 \text{U}(\theta)^\dagger - i \eta [\text{G},\text{U}(\theta)\rho_0\text{U}(\theta)^\dagger]+\mathcal{O}(\eta^2), \nonumber\\
&\mathcal{V}(\cdot) = -i [\text{G},\text{U}(\theta)(\cdot)\text{U}(\theta)^\dagger].\label{eq:pert_map} 
\end{align}
Then, the first order susceptibility of an eigenvalue $\varepsilon_\mu$ can be easily computed using perturbation theory: $\chi^{(1)} = \qqav{\text{l}_{\mu}|\mathcal{V}|\text{r}_{\mu}}$ (see App.~\ref{app:perturbation}). 

The analysis displayed in this section reveals the importance of coherent decay processes in preserving the subharmonic oscillations to long times. In the following, $\text{X}, \text{Y}$ and $\text{Z}$ denote the Pauli matrices.}

\subsection{A single qubit: Dephasing}\label{subsec:dephasing}

{\color{black} The simplest example of dissipative evolution supporting more than one steady state is \textit{pure dephasing}. Consider the Liouvillian
\begin{align}
\mathcal{L}(\rho) = -i (h/2) [\text{Z},\rho] + \kappa\left(\text{Z} \rho \text{Z}-\rho\right).
\label{eq:bistable_qubit}
\end{align} 
Since the Hamiltonian and the jumps commute we have $\text{dim}(\text{As}(\mathcal{H}))=2$. Introducing the basis $\text{Z}\ket{k} = (-)^k \ket{k}$,  we find the asymptotic states $\Psi_0 =\ketbra{0}{0}$ and $\Psi_1 = \ketbra{1}{1}$ both with eigenvalue $\lambda_\mu = 0$. This is a very particular evolution for which $\mathcal{L}^\ddagger = \mathcal{L}$ and, therefore, the two conserved quantities are $\text{j}_0 = \ketbra{0}{0}$ and $\text{j}_1 = \ketbra{1}{1}$. 
Alternatively, we can recover the trace as a conserved quantity by defining the rotated states $\tilde{\Psi}_\mu = (\Psi_0 + (-)^\mu\Psi_1)/2$, which leads to the conserved quantities $\text{j}_0 = 1$ and $\text{j}_1 = \text{Z}$. The subharmonic response is achieved by the unitary kick $\text{U}_\text{K}= \exp(-i \pi \text{X}/2) = -i \text{X}$}, and the Floquet propagator reads $\mathcal{E}_\text{F} = \text{X}\otimes\text{X} \exp(\mathcal{L} T)$. The spectrum of $\mathcal{E}_\text{F}$ is given by: 
\begin{align}
\{\varepsilon_\mu \} = \acom{1,-1,e^{-2 \kappa T},-e^{-2 \kappa T}}.
\end{align}
The basis of $\text{As}(\mathcal{H})_\text{F}$ is again given by $\tilde{\Psi}_\mu = \ketbra{0}{0} + (-)^\mu \ketbra{1}{1}$ fulfilling $\mathcal{E}_\text{F}(\tilde{\Psi}_\mu) = (-)^\mu \Psi_\mu$ with $\mu = 0,1$. Then, for an initial state of the form $\rho(0) = a \ket{0}\bra{0} + (1-a)\ket{1}\bra{1}$ with $a \in [0,1]$, we find $\rho(nT) =  a \ketbra{0\oplus n}{0\oplus n} + (1-a)\ketbra{1\oplus n}{1\oplus n}$, with $\oplus$ the sum modulo-2. Hence, it gives rise to subharmonic oscillations in the order parameter $\text{Z}$ of amplitude $2a -1$.

{\color{black} Finally, we compute the first order susceptibility $\chi^{(1)} =\qqav{\Psi_1|\mathcal{V}|\Psi_1}$. The perturbation map corresponding to a rotation error of the form $\text{U}_\text{K} = \exp(-i(\pi+\eta)\text{X}/2)$, is given by $\mathcal{V}(\cdot) = -i/2[\text{X},\mathcal{E}_\text{F}(\cdot)]$. For simplicity in the calculations, we take $h=0$ in Eq.~\eqref{eq:bistable_qubit}. The spectrum of $\mathcal{E}_\text{F}(\eta) = \mathcal{E}_\text{F}+\eta\mathcal{V}$ to first order in $\eta$ is 
\begin{align}
\{\varepsilon_\mu\} = \left\{1, -1, e^{-2 \kappa  T}, -e^{-2 \kappa T } \right\} + \mathcal{O}(\eta^2),
\end{align}}
which implies $\chi^{(1)}= 0$ indicating linearly robust subharmonic oscillations.

\subsection{Two qubits: decoherence free subspace}\label{subsec:dfs}
{\color{black}
We have studied a one-qubit scenario where $\text{As}(\mathcal{H})$ had two asymptotic states, corresponding to the diagonal elements of the density matrix. In this subsection, we consider the larger Hilbert space of two two level systems, $\mathcal{H} = \mathbb{C}^2\otimes \mathbb{C}^2$. {\color{black} A new structure of $\text{As}(\mathcal{H})$, that contains also coherences, can now arise.} These instances of asymptotic space are known as \textit{decoherence free subspaces} and have been studied in the literature as dissipation protected memories (see for instance \cite{Lidar1998,Beige2000,Barreiro2011,Albert2014}). For convenience, we also introduce the Bell basis: 
\begin{align}
\ket{\psi_\alpha} &= \frac{\ket{0,\alpha} + \ket{1,1\oplus \alpha}}{\sqrt{2}} \quad \& \quad \ket{\phi_\alpha} = \frac{\ket{0,\alpha} -\ket{1,1\oplus \alpha}}{\sqrt{2}}, \label{bell_basis}
\end{align}
for $\alpha = 0,1$ and $\oplus$ is sum modulo-2. 
We consider two different situations: \textit{independent} and \textit{collective} jump operators. After computing the asymptotic subspace and the conserved quantities of both scenarios, we also analyze the subharmonic oscillations induced by a kicked protocol.} \\

\textbf{Independent jump operators}\\
We consider first the case of no Hamiltonian and {\color{black} \textit{independent}} noise operators such that 
\begin{align}
\mathcal{L}_i(\rho) = \sum_\alpha \left( \text{L}^{\alpha}\rho \text{L}^{\alpha\dagger} - \frac{1}{2}\{\text{L}^{\alpha\dagger}\text{L}^\alpha,\rho\}\right) \quad \text{with} \quad \text{L}^\alpha = \ketbra{\psi_\alpha}{\phi_\alpha}.
\end{align}
{\color{black} In this context, independent refers to the fact that the jump operators contain a single transition.} Imposing $\mathcal{L}_i(\rho) =0$, one obtains $\text{As}(\mathcal{H}) = \text{span}\{ \Psi_{\alpha\beta} \}$ where $\Psi_{\alpha\beta} = \ket{\psi_\alpha}\bra{\psi_\beta}$ and, therefore, $\text{dim}(\text{As}(\mathcal{H})) = 4$. The conserved quantities read \cite{Albert2016}:
\begin{align}
\text{j}_{\alpha\beta} = \ketbra{\psi_\alpha}{\psi_\beta} + \delta_{\alpha\beta}\ketbra{\phi_\alpha}{\phi_\beta}.
\end{align}
{\color{black} The conserved quantities $\text{j}_{00}$ and $\text{j}_{11}$ indicate that the populations of the levels $\alpha = 0$ and $\alpha = 1$ are preserved. The other two conserved quantities $\text{j}_{01}$ and $\text{j}_{10}$ indicate that the coherence in the $\psi$-block is preserved, while the coherence in the $\phi$-block is destroyed by the dissipation.}\\

\textbf{Collective jump operator}\\ 
If now we consider a \textit{collective} jump operator such that
\begin{align}
\mathcal{L}_c(\rho) = \text{L} \rho \text{L}^\dagger - \frac{1}{2}\{ \text{L}^\dagger \text{L} ,\rho\} \quad \text{with} \quad \text{L} = \sum_\alpha \ketbra{\psi_\alpha}{\phi_\alpha}.
\end{align}
Imposing $\mathcal{L}_c(\rho) = 0$, the same asymptotic space $\text{As}(\mathcal{H}) = \text{span}\{ \Psi_{\alpha\beta} \}$ is found. The key difference is spotted by looking at the conserved quantities. We encounter 
\begin{align}
\text{j}_{\alpha\beta} = \ketbra{\psi_\alpha}{\psi_\beta} + \ketbra{\phi_\alpha}{\phi_\beta},
\end{align}
and therefore, the coherences within the $\phi$-block are preserved as well. {\color{black} As we shall see, the coherent preserving dissipation processes, provided by the collective jump operators, are crucial to preserve the subharmonic oscillations to later times.}

{\color{black} We introduce for convenience the \textit{local} Pauli matrices $\text{X}_r$, $\text{Y}_r$ and $\text{Z}_r$ for $r=1,2$ such that, for instance, $\text{X}_1\ket{00} = \ket{10}$. Likewise, we also define the total magnetization $\text{M}_a = \sum_{r} \text{A}_r$ for $a = x,y,z$. For both dissipative dynamics $\mathcal{L}_i$ and $\mathcal{L}_c$, subharmonic response is achieved with the unitary kick $\text{U}_\text{K} = \exp(-i (\pi/2) \text{M}_z)$. Using the same symbol $\mathcal{E}_\text{F} = \text{U}_\text{K}\otimes \text{U}_\text{K}^* \exp(\mathcal{L} T)$ for both $\mathcal{L} = \mathcal{L}_i$ and $\mathcal{L} =\mathcal{L}_c$, we obtain
\begin{align}
\mathcal{E}_\text{F}(\Psi_{\alpha\beta}) = (-)^{\alpha+\beta} \Psi_{\alpha\beta}.
\end{align}
Then, the coherence elements $\Psi_{01}$ and $\Psi_{10}$ display subharmonic response for $N=2$, while the populations are steady. Finally, we can compute the linear susceptibility $\chi^{(1)} = \qqav{\text{j}_{\alpha\beta}|\mathcal{V}|\Psi_{\alpha\beta}}$ with $\mathcal{V}(\cdot) = -(i/2)[\text{M}_z,\mathcal{E}_\text{F}(\cdot)]$. Using $\text{M}_z\ket{\psi_\alpha} = 2\delta_{\alpha 0}\ket{\phi_\alpha}$ it is easy to show that $\chi^{(1)}=0$ for, both, $\mathcal{L}_i$ and $\mathcal{L}_c$. Therefore, the system shows linearly robust subharmonic oscillations.}

{\color{black} A question remains still open. Is it possible to observe this time-crystalline behavior in a many-body open system, that is, where interactions play a role? Interestingly, the states $\ket{\psi_\alpha}$ can be interpreted as the two ground states of the Ising Hamiltonian of two particles $\text{H}_\text{S} = - J \text{X}_1\text{X}_2$. In the following sections, we investigate the many-body open system generalization of the Ising model.}

\section{The open XY model}\label{sec:openXY}

{\color{black} Let us investigate the stability of a time crystal in the paradigmatic  \textit{short-range} XY chain, described by the Hamiltonian:}
\begin{align}
\text{H}_\xi = - J\sum_{r=1}^L \left(\frac{1+\gamma}{2} \text{X}_r\text{X}_{r+1}+\frac{1-\gamma}{2} \text{Y}_r\text{Y}_{r+1} + h \text{Z}_r \right),
\end{align}
representing a 1D {\color{black}chain} of $L$ spins that interact anisotropically. {\color{black} This model has been largely investigated in the literature. For completeness, we summarize here its main features and show its quantum phase diagram in Fig.~\ref{fig:fig2a}}. We impose periodic boundary conditions ( i.e. $\text{X}_r = \text{X}_{L+r}$, and the similarly for $\text{Y}_r$, and $\text{Z}_r$), restrict ourselves to $L$ even, and gather the Hamiltonian parameters as $\xi = (J,\gamma,h)$. The Hamiltonian $\text{H}_\xi$ exhibits several symmetries (see, for instance, \cite{Franchini2017}): (\textit{i}) a rotation by $\pi/2$ along the z-axis interchanges the x and y spin interactions and it is equivalent to $\gamma \leftrightarrow -\gamma$, (\textit{ii}) a reflection of the spins with respect to the x-y plane is equivalent to $h \leftrightarrow -h$. Hence, it is sufficient to study the phase diagram for $\gamma, h \geq 0$. It is well known that this system undergoes a quantum phase transition at  $h=1$, that goes from an ordered phase for $h<1$ to a disordered phase for $h>1$. {\color{black} Moreover, the isotropic line $\gamma = 0$  displays a continuous symmetry generated by the magnetization $\text{M}_z = \sum_r \text{Z}_r$. The so-called Ising line, at $\gamma = 1$, separates the regions with attractive and repulsive interactions along the y-axis for $\gamma <1$ and $\gamma>1$ respectively. Finally, the ground state of a Hamiltonian lying on top of the line $h^2 +\gamma^2 = 1$ correspond to product ground states for all length chains $L$ with exact degeneracy in both parity blocks (see App.~\ref{app:product_gs} or \cite{Franchini2017}).}

{\color{black} A master equation of the form in Eq.~\eqref{eq:lindblad} arises from a microscopic system-reservoir description with: 
\begin{align}
&\text{H} = \text{H}_\xi + \text{H}_{SB} + \text{H}_{B}, \nonumber\\
&\text{H}_{SB} = {\color{black}\epsilon} \text{M}_z \otimes \text{B} \coloneqq  {\color{black}\epsilon} \sum_k \sum_{r=1}^L \text{Z}_r \otimes (g_{k} \text{b}_{k}^\dagger + g_{k}^* \text{b}_{k}) ,\nonumber\\
&\text{H}_{B} = \sum_{k}\Omega_{k} (\text{b}_{k}^\dagger \text{b}_{k} +1/2),\label{eq:system-bath}
\end{align}
where $k$ labels the different modes $\text{b}_k$ with frequency $\Omega_k>0$ of the bath, $g_k$ is the complex coupling strenght to that mode, and $\epsilon$ is a dimensionless perturbation parameter that we will set to one at the end of the computation. A similar system-bath Hamiltonian was used in \cite{Vogl2012} to study transport properties through spin chains. Note that the coupling to the bath is global in the sense that all particles are identically coupled to the reservoir $k$. In this section, we aim at deriving a dynamical equation for the reduced degrees of freedom of the system in two steps: first diagonalizing the system Hamiltonian $\text{H}_\xi$ and, second, tracing out the degrees of freedom of the bath. }

\subsection{Diagonalization of the XY model}\label{subsec:open_xy}

{\color{black} The first step towards the derivation of the master equation is to bring the Hamiltonian of the XY model into its diagonal form. It is well known that, for this model, this can be achieved using the Jordan-Wigner, Fourier and Bogoliubov transformations \cite{Lieb1961,Katsura1970,Franchini2017}. Note that $[\text{P},\text{H}_\xi] = 0$ where $\text{P} = \prod_r \text{Z}_r$ is the parity operator. Therefore, the diagonalization of $\text{H}_\xi$ can be done separately in two parity sectors with eigenvalues $p = \pm 1$. The well-known diagonalization procedure is shown by completeness in App.~\ref{app:diagonalization} and leads to:}
\begin{align}
\text{H}_\xi^{\pm} &= \frac{1}{2} \sum_{q \in \text{BZ}_\pm} E_{\xi,q} \left(\text{d}_{\xi, q}^\dagger \text{d}_{\xi, q}-\text{d}_{\xi, -q} \text{d}_{\xi, -q}^\dagger \right) = \sum_{q\in \text{BZ}_\pm} E_{\xi,q} (\text{d}_{\xi, q}^\dagger \text{d}_{\xi, q} -1/2),\label{eq:diagonal_ham}
\end{align}
where $\pm$ stands for the even and odd parity sectors, $\text{d}_{\xi,q}$ are the Bogoulibov fermions and the energy dispersion is given by:
\begin{align}
E_{\xi,q} = E_{\xi,-q} = 2 J \sqrt{\left(h-\cos\left(\frac{2\pi}{L}q\right)\right)^2+\left(\gamma \sin\left(\frac{2\pi}{L}q\right)\right)^2}. \label{eq:spectrum}
\end{align}
The label $q$ represents the quasi-momentum and takes values in the Brillouin zone:
\begin{align}
& \text{BZ}_+ = \{q = m +\frac{1}{2}\quad m \in \{-L/2, \cdots, L/2-1 \} \} & \text{for } \text{H}_\xi^+, \nonumber \\ 
& \text{BZ}_- = \{q =  \pm m  \quad m \in \{-L/2, \cdots, L/2-1 \} \} \,  & \text{for } \text{H}_\xi^- .
\end{align}
{\color{black}
Since $\omega_{\xi,q} >0$, the ground state $\ket{p,\text{GS}}$ of $\text{H}_\xi^{p}$ corresponds to the vacuum of Bogoulibov fermions in the $p$ parity sector.}

\subsection{Derivation of the master equation}

We are now ready to derive the master equation. We sketch here the crucial parts of the derivation, while details are given in App.~\ref{app:me_derivation}. For simplicity we assume only one reservoir, since the extension to multiple reservoirs is straightforward. The starting point is the {\color{black}well-known} Redfield equation in the rotating frame of $\text{H}_\xi + \text{H}_{B}$:
\begin{align}
\dot{\tilde{\rho}}(t) = - \int_0^\infty ds \text{tr}_{B}\com {\text{H}_{SB} (t), \com{ \text{H}_{SB} (t-s),\tilde{\rho}(t) \otimes \rho_\text{eq}}} + \mathcal{O}(\epsilon^3). \label{eq:second_order_main}
\end{align}
where $\rho_\text{eq} \propto \exp(-\beta \text{H}_B)$ is a Gibbs state at inverse temperature $\beta$. The jump operators arise from the decomposition of the $\text{M}_z$ into the eigenmodes of the system Hamiltonian. For every quasimomentum $q$, the magnetization can be divided into three different rotating frequencies: $\omega_{\xi,q}^\alpha \in \{ 0, 2 E_{\xi,q}, -2 E_{\xi,q} \}$ for $\alpha \in \{0,\uparrow, \downarrow\}$ respectively. {\color{black} Then, the magnetization can be decomposed as:
\begin{align}
&\text{M}_z(t) = -\sum_{q\in\text{BZ}_\pm} \sum_{\alpha} \text{L}_{\xi,q}^\alpha e^{i \omega_{\xi,q}^\alpha t},\label{eq:mz_partition}
\end{align}
where $\text{L}_{\xi,q}^0 = \cos\theta_{\xi,q} (\text{d}_{\xi,q}^\dagger \text{d}_{\xi,q}-\text{d}_{\xi,-q}\text{d}_{\xi,-q}^\dagger)$, $\text{L}_{\xi,q}^\uparrow = \sin\theta_{\xi,q} \text{d}_{\xi,q}^\dagger\text{d}_{\xi,-q}^\dagger  = \text{L}_{\xi,q}^{\downarrow \dagger}$, and the rotation angle 
\begin{align}
\theta_{\xi,q} = \tan^{-1}\left[\frac{\gamma \sin\left(\frac{2\pi}{L} q \right)}{h-\cos\left(\frac{2\pi}{L} q\right)}\right].
\end{align}}
We introduce the correlation function $C(t) = \text{tr}_\text{B}[\text{B}(t)\text{B}\rho_\text{eq}]$, and the real part of its Fourier transform $\kappa(\omega) = \text{Re}[C(\omega)]$. Inserting the decomposition in Eq.~\eqref{eq:mz_partition} and the definition of $\kappa(\omega)$ into Eq.~\eqref{eq:second_order_main}, it follows
\begin{align}
\dot{\tilde{\rho}}(t) = -{\color{black}\epsilon}^2\sum_{q, q'}\sum_{\alpha,\alpha'} \frac{\kappa(\omega_{\xi,q}^\alpha)}{2} e^{-i(\omega_{\xi,q}^\alpha - \omega_{\xi,q'}^{\alpha'})t } \left( \text{L}^{\alpha\dagger}_{\xi,q} \text{L}^{\alpha'}_{\xi,q'}\tilde{\rho}(t) -  \text{L}^{\alpha'}_{\xi,q'}\tilde{\rho}(t)\text{L}^{\alpha\dagger}_{\xi,q}\right) + \text{h.c.}.
\end{align}

The final step consists in using the secular approximation, that selects only those terms that fulfill the resonant condition $\omega^\alpha_{\xi,q} - \omega^{\alpha'}_{\xi,q'} = 0$. We now see that two different situations arise: in one hand, if $\xi$ is such that the dispersion of the energy as a function of $q$ is approximately flat, for instance at $(\gamma,h) = (1,0)$, the resonant condition results in $\alpha = \alpha'$. If, on the other hand, the dispersion is large enough, only those terms with $\alpha = \alpha'$ and $q=q'$ are resonant. These two conditions lead to collective and {\color{black} independent} decay processes in the sense of Subsec.~\ref{subsec:dfs}. {\color{black} Replacing $\epsilon\mapsto 1$, we arrive to the collective and independent Liouvillians:
\begin{align}
&\dot{\rho} = \mathcal{L}_c(\rho) = -i \com{\text{H}_\xi, \rho}+ \kappa^\downarrow_{\xi} \left( \text{L}_{\xi} \rho \text{L}^\dagger_{\xi} -\frac{1}{2}\acom{\text{L}^\dagger_{\xi} \text{L}_{\xi} , \rho}\right) + \kappa^\uparrow_{\xi} \left( \text{L}^\dagger_{\xi}\rho\text{L}_{\xi}-\frac{1}{2}\acom{\text{L}_{\xi} \text{L}^\dagger_{\xi}, \rho} \right),\label{eq:liouv_xy_coll}\\
&\dot{\rho} = \mathcal{L}_i(\rho) = -i \com{\text{H}_\xi, \rho}+ \sum_{q} \sum_{\alpha=\uparrow,\downarrow}\kappa_{\xi,q}^{\alpha} \left( \text{L}^\alpha_{\xi,q} \rho \text{L}^{\alpha\dagger}_{\xi,q} -\frac{1}{2}\acom{\text{L}^{\alpha\dagger}_{\xi,q} \text{L}^{\alpha}_{\xi,q} , \rho}\right).\label{eq:liouv_xy_loc}
\end{align}
where $\text{L}_\xi = \sum_q \text{L}^{\downarrow}_{\xi,q}$, $\kappa^{\uparrow\downarrow}_{\xi,q} = \kappa( \omega_{\xi,q}^{\uparrow\downarrow})$, and $\kappa^{\uparrow\downarrow}_{\xi} = \kappa( \omega_{\xi,0}^{\uparrow\downarrow})$. }
In analogy to the examples of Sec.~\ref{sec:intuition}, {\color{black} the nature of the jump operators determines whether the coherence is preserved in the decay process and, consequently, has a direct impact on the lifetime of the subharmonic oscillations.} 

\subsection{The battle against decoherence}\label{subsec:battle}

{\color{black} Before proceeding with the analysis of the open XY model as a time crystal, we discuss the decoherence process induced by $\mathcal{L}_c$ and $\mathcal{L}_i$.
First of all, for the open XY chain, the parity $\text{P}$ is a strong symmetry of the system, namely, it commutes with both, the Hamiltonian and the jump operators. In the vectorized picture, this strong symmetry leads to the partition:}
\begin{align}
\kket{\rho} = \begin{pmatrix}
\rho_{+}\\
\rho_\text{coh}\\
\rho_{-}\\
\end{pmatrix}, \qquad
\mathcal{L} = \left( \begin{array}{c|c|c}
\mathcal{L}_{+} & 0  & 0 \\ \hline
0 &  \mathcal{L}_\text{coh} & 0\\ \hline
0 & 0 & \mathcal{L}_-
\end{array}\right),
\end{align}
where $\mathcal{L}_\pm$ are \textit{bona fide} Liouvillians acting on the positive and negative parity blocks. It remains to check the action on the coherence part $\rho_\text{coh}$. {\color{black} For convenience, we introduce the notation $\ket{p,\vec{s}}$ for an eigenstate of parity $p$ of energy $E_{\vec{s}}$ and such that $\text{d}_{\xi,q}^\dagger \text{d}_{\xi,q} \ket{p,\vec{s}} = s_q$. Then, taking $\rho_\text{coh} = \ketbra{+,\vec{s}}{-,\vec{s}\, '}$ we have}
\begin{align}
\mathcal{L}_i \kket{\rho_\text{coh}}  = \left(-i (E_{\vec{s}} - E_{\vec{s}'}) - \sum_{q} \kappa^{\downarrow}_{\xi,q} \sin^2 (\theta_{\xi,q})(s_q+s'_q)\right)\kket{\rho_\text{coh}},
\end{align}
{\color{black} leading to an exponential decoherence process of all excited states. However, introducing $\Psi_{pp'} = \ketbra{p,\text{GS}}{p,\text{GS}}$, $\dot{\Psi}_{+-} = \dot{\Psi}_{-+} = 0$ when the two ground states are exactly degenerate. 

As we shall see, the subharmonic oscillations have a coherent origin. As we identified in Sec.~\ref{sec:intuition}, coherences can be preserved for collective decay processes. Hence, we expect systems with small dispersion, and therefore obeying the collective equation $\dot{\rho} = \mathcal{L}_c(\rho)$, to display more rigid oscillations than those with a large dispersion well-described by $\dot{\rho} = \mathcal{L}_i(\rho)$. } 

\section{Characterization of the open XY time crystal}\label{sec:results}

{\color{black} Having derived the master equation of the open XY model from a microscopic description, we aim now at characterizing the behavior of the system under  periodic driving. To take advantage of the knowledge on the quantum phase diagram obtained in Subsec.~\ref{subsec:open_xy}, we restrict ourselves to study the time crystal also at zero temperature. In this scenario, we find a multidimensional $\text{As}(\mathcal{H})$ that supports time-crystalline order. We consider the kicked protocol consisting in the dissipative evolution of the open XY chain during a time $T$, followed by the instantaneous kick $\text{U}_\text{K} = \exp\left(-i \pi \text{M}_z/2\right) = \text{P}.$ We are primarily interested in evaluating properties (I)-(III) stated in Sec.~\ref{sec:time-crystals}. Our analysis relies on exact diagonalization of the Liouvillian matrix, which becomes costly very fast, as the number of spins $L$ increases. Recall that the dimension of $\mathcal{L} \in \text{Op}(\mathcal{H}\otimes\mathcal{H})$ is $\text{d}_\mathcal{H}^2\times \text{d}_\mathcal{H}^2$. For this reason, we analyze the dynamics for relative small system size under the use of periodic boundary conditions, reducing then the effects arising from the finite size of the lattice.

As discussed previously, the rigidity must be defined with respect a given perturbation. Throughout this section, we consider a global rotation error $\eta$ in the rotation angle such that
\begin{align}
\text{U}_\text{K} = \exp\left(-i (\pi +\eta)\frac{\text{M}_z}{2}\right).\label{eq:error_kick}
\end{align}
For this error, the perturbation map has the form $\mathcal{V}(\cdot) = -(i/2) [\text{M}_z,\mathcal{E}_\text{F}(\cdot)]$. Moreover, in Subsec.~\ref{subsec:disorder} we briefly discuss the case where the error depends on the chain site which induces disorder in the system.}

\subsection{The phase diagram}

{\color{black}
It is expected, from the analysis in Subsec.~\ref{subsec:battle}, that the subharmonic oscillations are more robust at those points of the diagram where collective decay processes are predicted. Recall that collective decay occurs when the dispersion of the energy is approximately flat. Here, we aim at quantifying the robustness through the dissipative Floquet gap $\Delta_\text{F}$. As discussed in Subsec.~\ref{subsec:observations}, the kicked protocol considered relies on the degeneracy of the ground states $\ket{\pm,\text{GS}}$. If the degeneracy is not exact, the subharmonic oscillations show a beating behavior that destroys the periodic structure of the oscillations. However, in the thermodynamic limit the degeneracy condition is guaranteed for any point in the ordered phase ($h<1$). For finite systems, as in our numerical study, the exact degeneracy is only found along the factorization line $\gamma^2+h^2=1$. Along this line, the dispersion is given by 
\begin{align}
\omega_{\xi,q} = J\left|1-h\cos\left(\frac{2\pi}{L}q\right)\right|,
\end{align} 
and therefore, we expect the sub-harmonic oscillations to be more robust against errors for small magnetic field $h$. In Fig.~\ref{fig:fig2b}, we show a contour plot of the dissipative Floquet gap $\Delta_\text{F}$ as a function of the transverse field $h$ and the error $\eta$ in Eq.~\eqref{eq:error_kick}. Our calculations are done at the factorization line for a chain of $L=6$ spins. The inverse Floquet gap $\Delta^{-1}_\text{F}$ fixes the time-scale of decay of the oscillations and, therefore, it is a measure of the rigidity of the oscillations. As expected, we see in Fig.~\ref{fig:fig2b}, that the rigidity of the oscillations decreases as we turn on the transverse field $h$. After a value $h\approx0.2$, the rigidity of the oscillations stabilize. Finally, at a transverse field $h\approx 0.9$ the region of rigid oscillations increases again. This behavior is explained the following way: in the limit $h\to 1$, the Hamiltonian $\text{H}_\xi$ commutes with the magnetization $\text{M}_z$ implying that they have a common eigenbasis. Then, coherence element $\Psi_{+-} = \ketbra{+,\text{GS}}{-,\text{GS}}$ is a right eigenvector of $\mathcal{E}_\text{F}(\eta)$ and has, for all values of $\eta$, a modulus one eigenvalue. Therefore, the oscillations do not decay and $\Delta_\text{F}\to 0$. However, this rises an important remark: there are systems, which have non-decaying oscillations that are not necessarily subharmonic. In the case $h \to 1$, the non-decaying oscillations show a beating pattern that breaks the subharmonic response. For this reason, in Fig.~\ref{fig:fig2c} we show a complementary contour plot of the magnitude $\delta_\text{F} = |\varepsilon_\star+1|$, namely the distance to the time-crystal state.  The quantity $\delta_\text{F}$ indicates how faithful are the subharmonic oscillations and, for this reason, we may look at Fig.~\ref{fig:fig2c} as a \textit{pseudo} phase diagram for the time crystal phase. In Fig.~\ref{fig:fig2c}, the stabilizing effect of the coherent decay processes is clearer as the region of stability diminishes clearly when $h$ increases. Finally, the linear susceptibility $\chi^{(1)} = |\partial \varepsilon_\star /\partial \eta|_{\eta=0}$ can be identified as 
\begin{align}
\chi^{(1)} = \lim_{\eta \to 0} \frac{|\varepsilon_\star - (-1)|}{\eta} =  \lim_{\eta \to 0}  \left| \frac{\delta_\text{F}}{\eta}\right|,
\end{align}
and, therefore, can be read from Fig.~\ref{fig:fig2c}.}

\subsection{Robustness to disorder}\label{subsec:disorder}
{\color{black}
At the time of completion of this manuscript it was noted in \cite{Tindall2019} that a permutationally invariant system is linearly robust to disorder. A similar idea applies to the open-XY time crystal. To be more precise, consider that we kick the system with the disordered pulse
\begin{align}
\text{H}_\text{S}'(t) = \sum_{n} \delta (t-nT) \sum_{r=1}^{L} (\pi + \eta \delta_r )\frac{\text{Z}_r}{2},
\end{align}
where $\eta$ parametrizes the disorder strength, and $\delta_r$ measures of the fluctuation at site $r$ subject to $\sum_r \delta_r = 0$. To first order, the disordered perturbation map $\mathcal{V}_\delta$ is
\begin{align}
\mathcal{V}_\delta(\rho) = - i\frac{\eta}{2}\sum_{r}\delta_{r} [\text{Z}_r,\mathcal{E}_\text{F}(\rho)].
\end{align}
Hence, the first order susceptibility $\chi^{(1)}$ can be computed as
\begin{align}
\chi^{(1)}_\mu = \frac{|\varepsilon_\mu|}{2} \left|\qqav{\text{l}_\mu|\text{Z}_1\otimes \text{1}-\text{1}\otimes \text{Z}_1 |\text{r}_\mu}\right| \left(\sum_{r}\delta_{r}\right) = 0
\end{align}
where we have used the permutationally invariance of the system. Therefore, it is guaranteed that the system is linearly robust to disorder.}

\subsection{Time translation symmetry breaking}

\begin{figure}
\sidesubfloat[]{\includegraphics[width=0.43\textwidth]{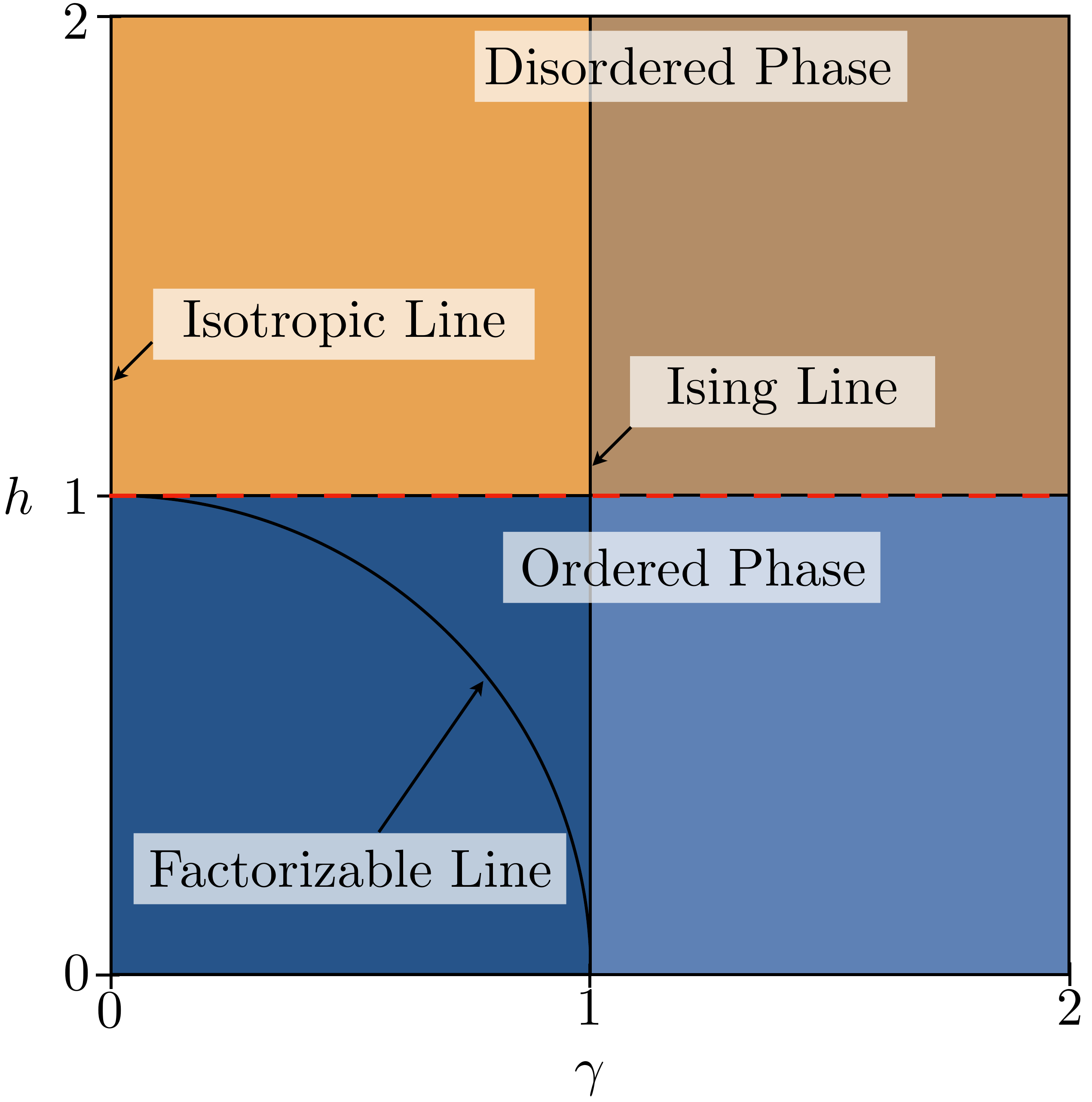}\label{fig:fig2a}}\\
\sidesubfloat[]{\includegraphics[width=0.49\textwidth]{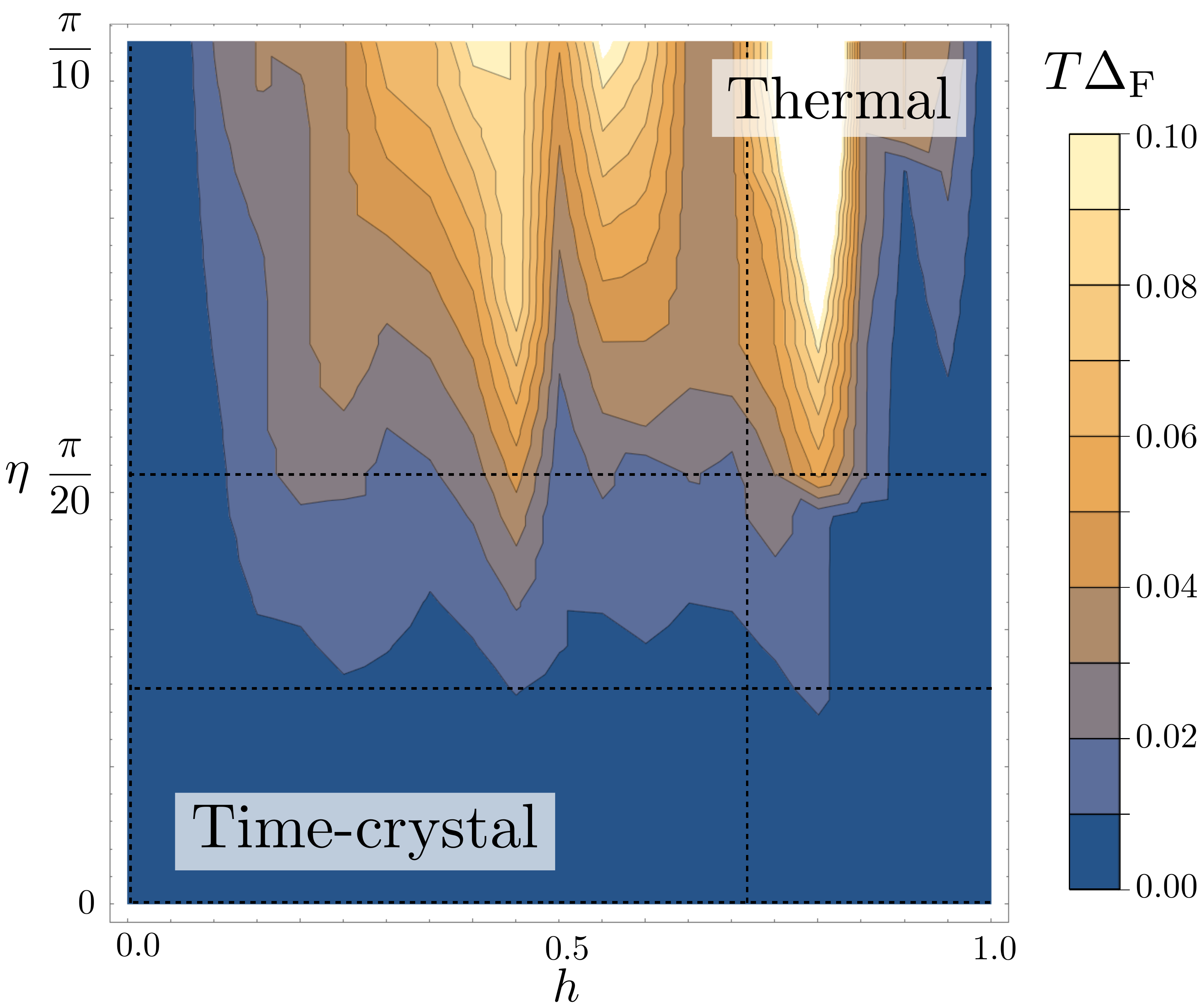}\label{fig:fig2b}}
\sidesubfloat[]{\includegraphics[width=0.49\textwidth]{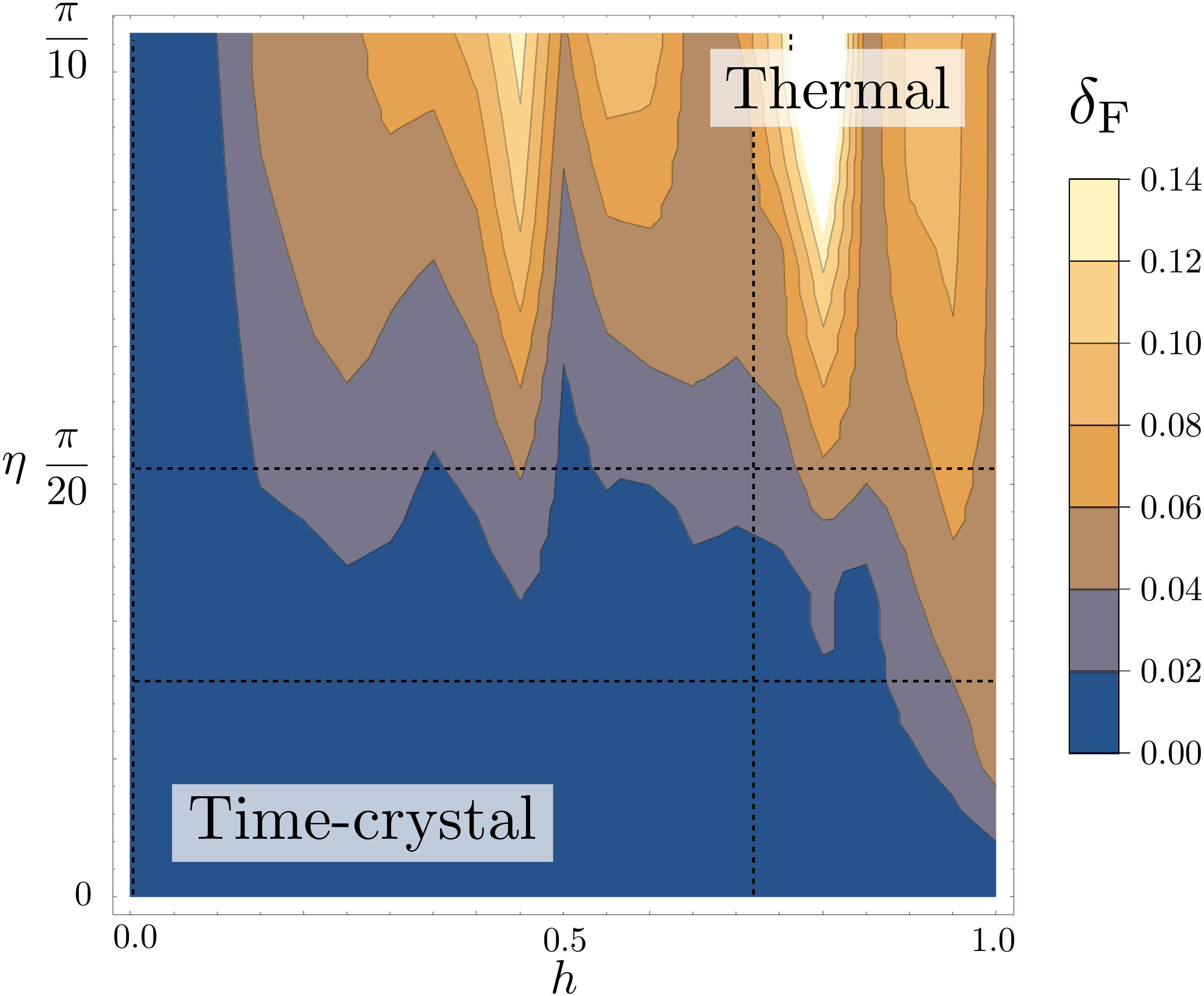}\label{fig:fig2c}}
\caption{\color{black} (a) The quantum phase diagram of the XY model with the quantum phase transition $h=1$ represented by the red dashed line. (b) Contour plot of the dissipative Floquet gap $\Delta_\text{F}$ along the factorization line $h^2 + \gamma^2 =1$ for $L=6$ spins. The vertical dashed lines correspond to $h=0, 1/\sqrt{2}$ while the horizontal dashed lines correspond to $\eta = \pi/40, \pi/20$. (c) Contour plot of $\delta_\text{F} = |\varepsilon_\star+1|$ along the factorization line $h^2 + \gamma^2 =1$ for $L=6$ spins. The vertical dashed lines correspond to $h=0, 1/\sqrt{2}$ while the horizontal dashed lines correspond to $\eta = \pi/40, \pi/20$.\\
Parameters: $JT = 10$, $\beta^{-1} = 0$, $\kappa(\omega) = \kappa_0 \omega$, $\kappa_0 = 0.01 J$ \label{fig:fig2}.}
\end{figure}

\begin{figure}
\sidesubfloat[]{\includegraphics[width=0.44\textwidth]{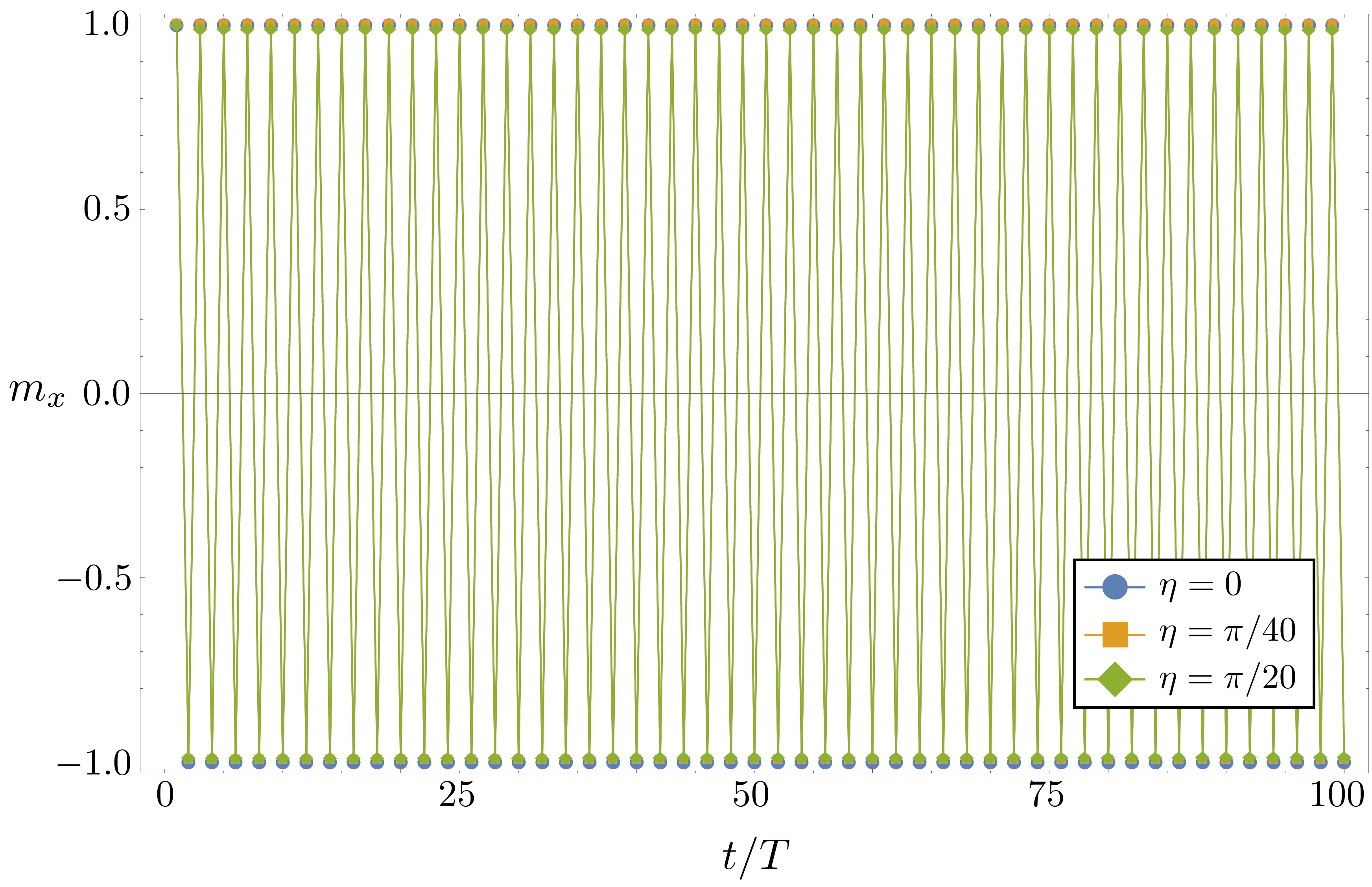}\label{fig:fig3a}}\quad
\sidesubfloat[]{\includegraphics[width=0.44\textwidth]{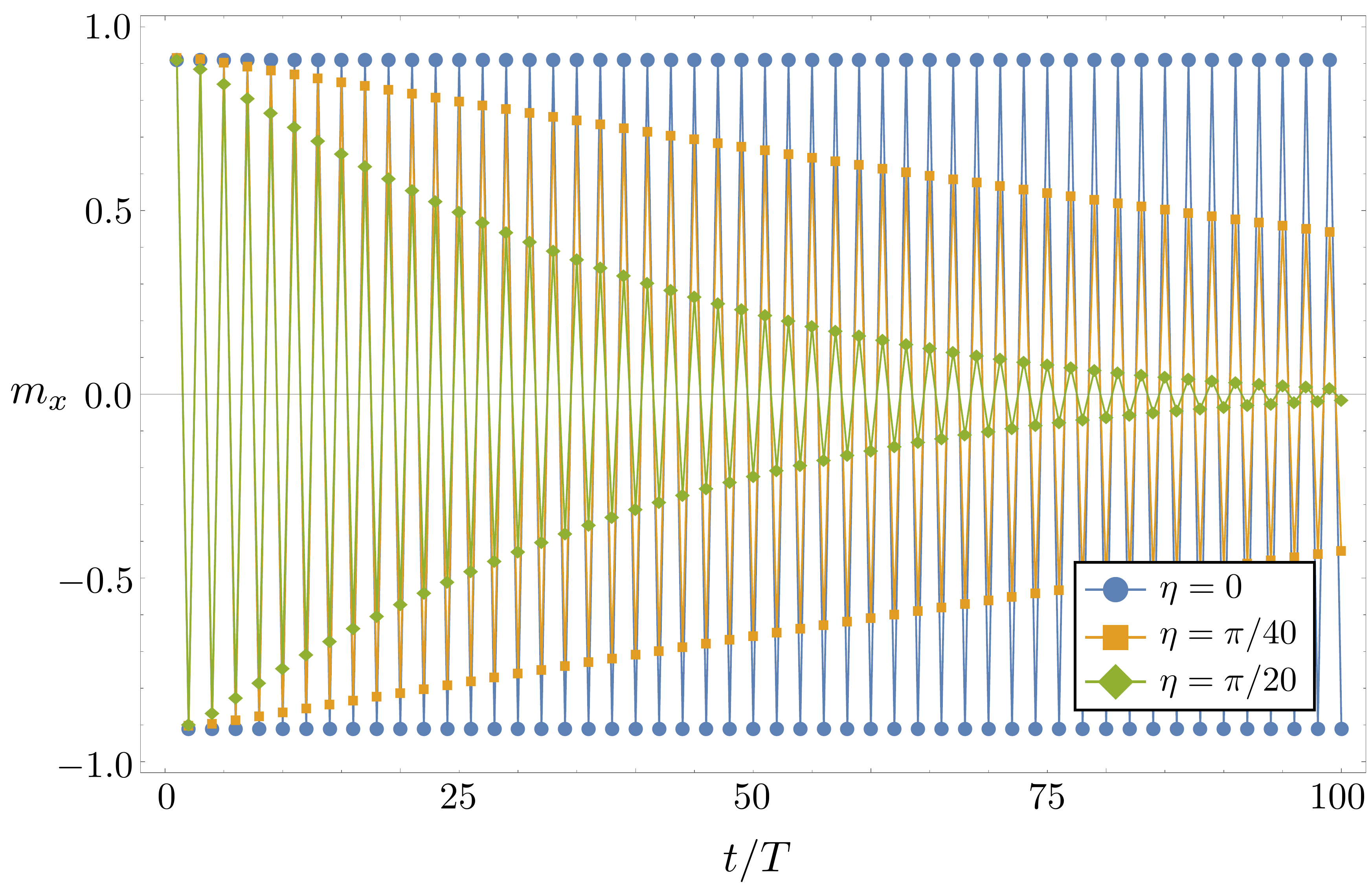}\label{fig:fig3b}}\\
\sidesubfloat[]{\includegraphics[width=0.44\textwidth]{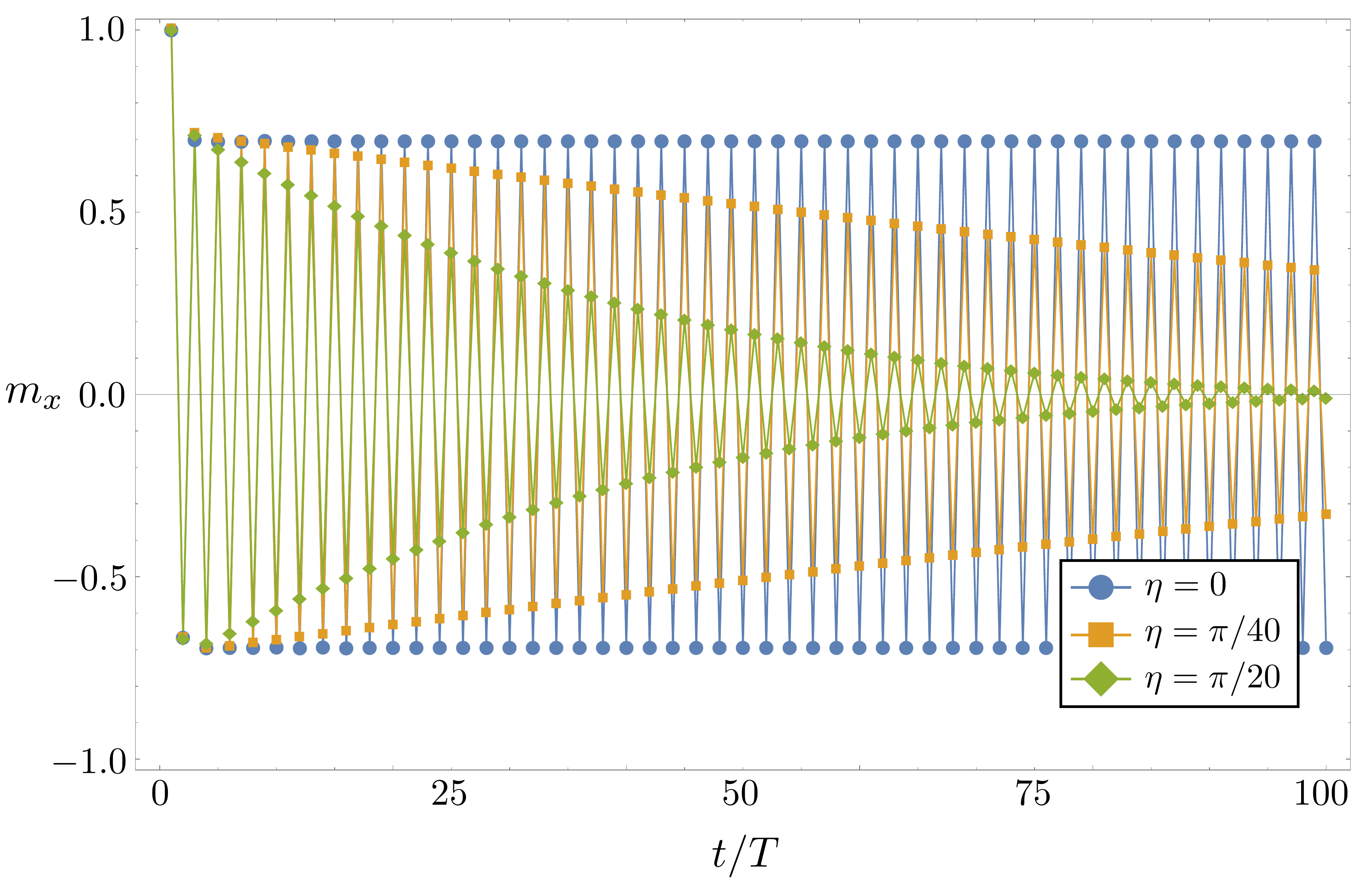}\label{fig:fig3c}}\quad
\sidesubfloat[]{\includegraphics[width=0.44\textwidth]{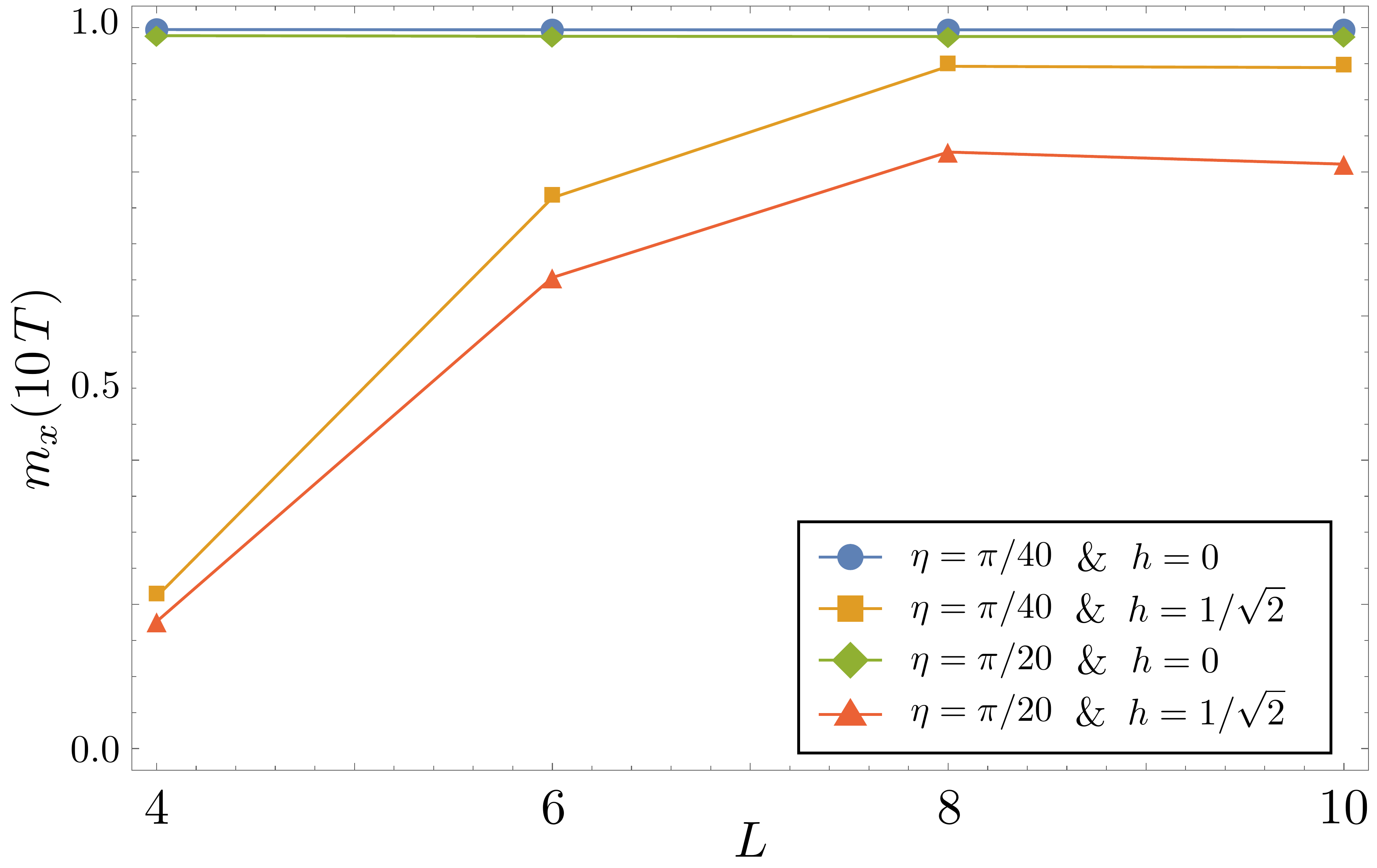}\label{fig:fig3d}}\\
\caption{\color{black}{(color-online) (a) Subharmonic oscillations for a chain of length $L$ at the Ising point $(\gamma,h) = (1,0)$. (b) Subharmonic oscillations for a chain of length $L$ at $(\gamma,h) = (1/\sqrt{2},1/\sqrt{2})$. (c) Subharmonic oscillations for a chain of length $L$ at $(\gamma,h) = (1/\sqrt{2},1/\sqrt{2})$  with a local initial state. (d) Scaling with the chain length $L$ of the amplitude of the subharmonic oscillations after 10 periods.
Parameters: $JT = 10$, $\beta^{-1} = 0$, $\kappa(\omega) = \kappa_0 \omega$, $\kappa_0 = 0.01 J$. \label{fig:fig3}}}
\end{figure}

{\color{black} The most characteristic feature of time crystals is discrete time-translation symmetry breaking. For the open XY time-crystal, a valid order parameter is the  \textit{local} observable $\text{m}_x = 1/L \sum_r \text{X}_r$, namely, the magnetization per spin along the x-axis. This observable does not transform trivially under the parity operation and, therefore, it detects the subharmonic oscillations. In Fig.~\ref{fig:fig3a} and Fig.~\ref{fig:fig3b} we show the subharmonic oscillations for chain of $L=6$ spins with parameters $(\gamma,h) = (1,0)$ and $(\gamma,h) = (1/\sqrt{2},1/\sqrt{2})$ respectively. For each parameter set, we consider rotation errors $\eta = (0,\pi/40,\pi/20)$ which corresponds to the crossings of the dotted lines in Fig.~\ref{fig:fig2b}. As the initial state of the evolution we take $\ket{\psi(0)} = (\ket{+,\text{GS}}+\ket{-,\text{GS}})/\sqrt{2}$. In Fig.~\ref{fig:fig3a}, we observe that the oscillations are rigid against errors for $(\gamma,h) = (1,0)$ where coherent decay processes are expected.
Contrarily, in Fig.~\ref{fig:fig3b} we observe that the oscillations decay faster when we introduce the rotation error $\eta$, in agreement with the computation of $\Delta_\text{F}$ in Fig.~\ref{fig:fig2b}. Another relevant observation is that the amplitude of the oscillations decreases with $h$. The amplitude of the oscillations can be analytically computed for $\eta = 0$ along the factorization line (see App.~\ref{app:product_gs}) and is given by:
\begin{align}
|m_x(nT)| = \sqrt{\frac{2\gamma}{1+\gamma}} = \sqrt{\frac{2\sqrt{1-h^2}}{1+\sqrt{1-h^2}}}.
\end{align}  
It is important to note that the oscillations can also be seen for an initially \textit{local} state. In Fig.~\ref{fig:fig3c}, we show the subharmonic oscillations at $(\gamma,h) = (1/\sqrt{2},1/\sqrt{2})$ for the initial pure state $\ket{\psi(0)} = \bigotimes_r (\ket{0}_r +\ket{1}_r)/\sqrt{2}$. We observe that the subharmonic oscillations reappear after a transient time. However, due to the incoherent decay processes, the amplitude is slightly damped.}

\floatsetup[figure]{style=plain,subcapbesideposition=top}

\floatsetup[figure]{style=plain,subcapbesideposition=top}

\subsection{Scaling with the length of the chain}

{\color{black} In this subsection, we aim at studying the scaling of the time-crystalline behavior with the chain length $L$. As we have briefly discussed, we expect the finite size effects of a closed chain with periodic boundary conditions to become quickly unimportant. Exact diagonalization of a Liouvillian of $L=8$ spins is already computationally demanding. Therefore, we avoid using this method to study the robustness of the oscillations. However, we are still able to integrate the equation of motion using, for instance, a standard Runge-Kutta method. In Fig.~\ref{fig:fig3d}, we show the amplitude of the oscillations in the magnetization after ten periods $m_x(10 T)$ as a function of the chain length $L$, that we are able to scale up to $L = 10$ spins. Fixing the error of the protocol $\eta$, we observe that, as $L$ increases, the system becomes more robust to the same error of rotation, and the amplitude $m_x(10T)$ becomes larger. However, already for $L = 8$ particles, it stabilizes as finite size effects are no longer important. This result indicates that the subharmonic oscillations are both, present and robust, in a wide range of lengths that go from relatively small system sizes to the thermodynamic limit.}

\section{Conclusions}\label{sec:conclusions}

In this article we have presented some self-contained results concerning the existence and properties of time crystals in open systems whose evolution is described with a Lindblad master equation. After introducing the tools of Markovian quantum open system dynamics, we have provided a compact definition of an open system time crystal derived from the properties the spectrum of its Floquet propagator. We have, as well, identified which are the most relevant properties of this object with special emphasis on the asymptotic subspace and the associated conserved quantities. We have analytically solved the kicked dynamics of an exemplary set of one and two-qubit open system models and exploit such analysis to provide key features on the properties and stability of time crystals in open systems.
Finally, we have derived and analyzed the {\color{black}\textit{short-range}} open XY model as a time-crystal. There has been some discussion around the possibility that only collective models can exhibit time-crystalline order in open quantum systems, our analysis shows that this is not the case and we conclude that long-range interactions are not crucial features to observe time-crystalline behavior. Nonetheless, our findings show that collective jump operators are crucial in order to have subharmonic oscillations that are more robust to rotation errors. Intuitively, the collective jump operators help to preserve coherence in the dissipation process, and the time-crystalline oscillations are usually coherent in the Hamiltonian eigenbasis. Moreover, at the time of completion of this manuscript, it was noted in \cite{Tindall2019} that permutationally invariant systems are, also, robust to disorder. In Subsec.~\ref{subsec:disorder} we show that the same idea applies to the open XY time-crystal presented in this work.

To conclude, in agreement with \cite{Lazarides2019}, we believe that a promising direction of investigation is that of non-Markovian environments. There, the backflow of information to the system may be controlled to achieve subharmonic response.

\section*{Aknowledgements}

We thank Philipp Strasberg, Michalis Skotiniotis and Giacomo Guarnieri for fruitful discussions. We acknowledge support from the Spanish MINECO, project FIS2016-80681-P, and from the Catalan Government: projects CIRIT 2017-SGR-1127, AGAUR FI-2018-B01134, and QuantumCAT 001-P-001644 (RIS3CAT comunitats), co-financed by the European Regional Development Fund (FEDER).
\newpage
\appendix

\section{Mathematical properties of $\mathcal{L}$}\label{app:generator_math}

For completeness, we include here some discussion about the mathematical properties outlined in Sec.~\ref{sec:preliminaries} of the main text.

If a particular eigenvalue $\lambda_\mu$ of $\mathcal{L}$ has algebraic multiplicity $m_\mu\geq1$, the number of non-trivial solutions of Eq.~\eqref{eq:right_evec} lies between one and $m_\mu$. If there is strictly one solution $\kket{\text{r}_\mu}$ associated to $\lambda_\mu$ but $m_\mu >1$, higher rank generalized eigenvalues can be found as solutions of the recursive equation $(\mathcal{L}-\lambda_\mu\mathcal{I}) \kket{\text{r}_\mu(s)} = \kket{\text{r}_\mu(s-1)}$, where $\kket{\text{r}_\mu(1)} = \kket{\text{r}_\mu}$ and $s$ denotes the rank. Note that $(\mathcal{L}-\lambda_\mu\mathcal{I})^k \kket{\text{r}_\mu(s)}=0$ only if $k\geq s$.
\begin{enumerate}
\item[(i)] \textbf{Spectrum of the Liouvillian:}
\end{enumerate}
Consider $\text{r}_\mu \in \text{Op}(\mathcal{H})$ such that $\mathcal{L}(\text{r}_\mu) = \lambda_\mu \text{r}_\mu$. Then, hermiticity preservation 
$\mathcal{L}(\text{r}_\mu ^\dagger) = (\mathcal{L}(\text{r}_\mu))^\dagger$ guarantees
\begin{align}
\mathcal{L} (\text{r}_\mu^\dagger) = (\mathcal{L}(\text{r}_\mu))^\dagger = \lambda_\mu^* \text{r}_\mu^\dagger,
\end{align}
that is, either the eigenvalues are real or come by conjugate pairs. Note that if $\text{r}_\mu = \text{r}_\mu^\dagger$ $\lambda_\mu \in \mathbb{R}$. The converse is true, at least, when $\lambda_\mu$ is non-degenerate. 
\begin{enumerate}
\item[(ii.1)] \textbf{Eigenvectors of different eigenvalue are linearly independent:}
\end{enumerate}
Consider the linear combination $\sum_{\mu} c_\mu \kket{\text{r}_\mu} = 0$, where $\kket{\text{r}_\mu}$ are eigenvectors of different eigenvalue of $\mathcal{L}$. If one multiplies the linear combination by the operator $\prod_{\nu\neq\mu}(\mathcal{L}-\lambda_\nu\mathcal{I})$ it leads to $c_\mu \text{r}_\mu = 0$, and therefore the set $\{\kket{\text{r}_\mu}\}_\mu$ is linearly independent by definition. 
\begin{enumerate}
\item[(ii.2)] \textbf{Generalized eigenvectors are linearly independent:}
\end{enumerate}
Consider the simplified case where there is only one eigenvector $\kket{\text{r}_\mu(0)}$ for the eigenvalue $\lambda_\mu$. Consider the linear combination $\sum_k c_k \kket{\text{r}_\mu(k)} = 0$. Then, by successive applications of the operator $\mathcal{L}-\lambda_\mu\mathcal{I}$ we obtain $c_k = 0$ $\forall k$ and then the set $\{\kket{\text{r}_\mu(k)} \}_k$ is linearly independent.

This prove can be extended using similar methods to the complete set of generalized eigenvectors $\{\kket{\text{r}_\mu(k)}\}_{\mu,k}$ of $\mathcal{L}$. Then, the complete set of generalized eigenvectors forms a basis of the space.

\begin{enumerate}
\item[(iii.1)] \textbf{Biorthogonality of left and right eigenvectors:}
\end{enumerate}
Consider $\{\kket{\text{l}_\mu(k)} \}_{\mu,k}$ and $\{\kket{\text{r}_\mu (k)}\}_{\mu,k}$ the set of left and right generalized eigenvectors. Then,
\begin{align}
\qqav{\text{l}_\mu(1)|(\mathcal{L}-\lambda_\nu\mathcal{I})|\text{r}_\nu (1)} = 0 \Rightarrow (\lambda_\mu - \lambda_\nu) \qqav{\text{l}_\mu(1)|\text{r}_\nu(1)}=0,
\end{align}
and, therefore, eigenvectors of different eigenvalues can be chosen biorthonormal. For a diagonalizable matrix, the biorthogonal relation can be compactly written as $\mathcal{W}_l^\ddagger \mathcal{W}_r = \mathcal{I}$.

\begin{enumerate}
\item[(iii.2)] \textbf{Normal Jordan form:}
\end{enumerate}
Given $\mathcal{L}$, it exists a similarity transformation $\mathcal{W}_r$ such that $\mathcal{W}_r^{-1} \mathcal{L} \mathcal{W}_r = \mathcal{J}$ where $\mathcal{J}$ is in Jordan canonical form such that the columns $\mathcal{W}_r$ are the generalized eigenvectors $\kket{\text{r}_\mu(s)}$. This corresponds to solving the generalized eigenvalue equation $\mathcal{L} \mathcal{W}_r = \mathcal{W}_r \mathcal{J}$.

\begin{enumerate}
\item[(iv)] \textbf{Time-ordered propagator:}
\end{enumerate}
{\color{black} In the more general case where the Liouvillian $\mathcal{L} = \mathcal{L}(t)$ depends on time, the evolution map $\mathcal{E}(t)$ is given by the well-known solution 
\begin{align}
\mathcal{E}(t) = \mathcal{T}\exp\left(\int_0^{t} \mathcal{L}(s) ds \right),
\end{align}
where $\mathcal{T}$ denotes the time-ordering operator, which arises because the Liouvillian matrix may not commute with itself at different times. Then, the steady-state is, in general, time-dependent. However, since the trace is preserved, the left-eigenvector is $\llangle{1}|$ at all times. }

\begin{enumerate}
\item[(v.1)] \textbf{Existence of the steady-state:}
\end{enumerate}
The trace preserving condition for an arbitrary state $\rho$, together with Eq.~\eqref{eq:adjoint_lindblad} lead to:
\begin{align}
\tr{ \mathcal{L}(\rho)} = \tr{(\mathcal{L}^\ddagger(\text{1}) )^\dagger \rho}=0,
\end{align}
guarantees at least one eigenvalue $\lambda_0 = 0$. The corresponding eigenvector $\text{r}_0$ fulfills that $\partial_t \text{r}_0 = 0$, and it is often referred to the steady-state. In general, however, the steady-state may not be unique.

\begin{enumerate}
\item[(v.2)] \textbf{Contractivity of the evolution. Convergence to $\textnormal{As}(\mathcal{H})$:}
\end{enumerate}
We include it here the proof given in \cite{Ruskai1994}. Given an Hermitian operator $\text{A} = \text{A}_+ - \text{A}_-$ where $\text{A}_\pm$ are positive matrices, and a CPTP map $\mathcal{E}$, it follows that 
\begin{align}
\text{tr}|\mathcal{E} (\text{A})| &= \text{tr}|\mathcal{E} (\text{A}_+) -  \mathcal{E} (\text{A}_-)| \leq \text{tr}|\mathcal{E} (\text{A}_+)| + \text{tr}|\mathcal{E}(\text{A}_-)| = \text{tr}|\text{A}_+|+ \text{tr}|\text{A}_-| = \text{tr}|\text{A}|.
\end{align}
In particular, given $\rho,\sigma \in \text{S}(\mathcal{H})$ we see that the trace distance $\text{D}(\mathcal{E}(\rho),\mathcal{E}(\sigma)) \leq \text{D}(\rho,\sigma)$, which provides a convergence towards the asymptotic subspace. 

\section{Proofs of the observations}\label{app:proofs}

Here we gather the proofs of the observations in Subsec.~\ref{subsec:observations}.

\textbf{Proof of observation 1}\\
It follows from Eq.~\eqref{floquet_propagator} the form of the Floquet propagator. Notice that
\begin{align}
\bbra{\text{1}}\mathcal{E}_\text{F}
  = \left( \exp(\mathcal{L}^\ddagger T) \text{U}^\dagger_\text{K} \otimes \text{U}_\text{K}^\text{T} \kket{\text{1}} \right)^\ddagger =  \left(\exp(\mathcal{L}^\ddagger T) \kket{\text{1}})\right)^\ddagger = \bbra{\text{1}},
\end{align}
and therefore $\bbra{\text{1}}\mathcal{E}_\text{F} = \bbra{\text{1}}$ has eigenvalue 1. Alternatively, it is a consequence of the fact that a concatenation of trace preserving maps is also trace preserving.\\

\textbf{Proof of observation 2}\\
The characteristic equation is, in general, not invariant under \textit{one-sided} unitary transformations, i.e., $P_\mathcal{E}(\lambda) = {\color{black}\text{det}}(\mathcal{E}-\lambda \mathcal{I}) \neq \text{det}(\mathcal{E}'-\lambda \mathcal{I}) = P_{\mathcal{E}'}(\lambda) $.\\

\textbf{Proof of observation 3}\\
It follows from noting that the characteristic equation is invariant under unitary transformations, i.e., $P_\mathcal{E}(\lambda) = {\color{black}\text{det}}(\mathcal{E}-\lambda \mathcal{I}) = \text{det}(\mathcal{E}''-\lambda \mathcal{I}) = P_{\mathcal{E}''}(\lambda) $.\\

\textbf{Proof of observation 4}\\
By contradiction, if it exists one and only one eigenstate $\Psi_0$ with $|\varepsilon_0| = 1$, it corresponds to a normalizable state with $\text{tr}[\Psi_0]=1$. Given any initial state $\rho(0)$
\begin{align}
\lim_{n\to\infty} \rho(n T) = \lim_{n \to\infty} \mathcal{E}_\text{F}^n (\rho(0)) = \Psi_0,
\end{align}
which proves that, asymptotically, the state has the same periodicity than the Liouvillian, i.e., $\rho(t_n+T) = \rho(t_n)$. The converse implication fails, for instance, if there are two non-decaying states with the same eigenvalue $\varepsilon_{0,1} = \varepsilon_{0,2}$,  one finds a $T$-periodic response even though $\text{dim As}(\mathcal{H}) > 1$.\\

\textbf{Proof of observation 5}\\
Consider the unitary kick $\text{U}_\text{K} \coloneqq \sum_k \ketbra{\psi_k} {\phi_k}$. Then all the elements of $S_u$ fulfill $\mathcal{E}_\text{F}(\ketbra{\psi_k}{\psi_{k'}}) = \ketbra{\psi_k}{\psi_{k'}}$ and are therefore part of $\text{As}(\mathcal{H})$ of the map $\mathcal{E}_\text{F}$. {\color{black} Note that the positivity of $\mathcal{E}$ only allows off-diagonal elements of the form $\ketbra{\psi_k}{\psi_{k'}}$ in $S_u$ if also $\ketbra{\psi_k}{\psi_k}$ and $\ketbra{\psi_k}{\psi_{k'}}$ belong to $S_u$.}\\

\textbf{Proof of observation 6}\\
Let us define $\{\Psi_\mu \}$ a basis of $\text{As}(\mathcal{H})$ of $\mathcal{E}$. Consider the Floquet map $\mathcal{E}_\text{F} = \text{U}_\text{K}\otimes\text{U}_\text{K}^* \mathcal{E}$. From condition (i) follows
$\mathcal{E}_\text{F} (\Psi_\mu) = \mathcal{U}_\text{K} (\Psi_\mu)$.
Condition (ii) implies that:
\begin{align}
\rho(nT) = \sum_{\mu} j_\mu \mathcal{U}_\text{As}^n (\Psi_\mu) + \mathcal{O}(e^{-\Delta n T}),
\end{align}
where $\text{j}_\mu$ is the associated conserved quantity to $\Psi_\mu$, and $j_\mu =\text{tr}[\text{j}_\mu^\dagger \rho(0)] $. Finally, for large enough $m$, condition (iii) implies that exists an $\text{O}$ such that $O(T)\neq O(0) = O(NT)$. 

\section{Generalized susceptibilities and higher order robustness}\label{app:perturbation}

In the main text we considered the linear susceptibility $\chi^{(1)}$. However, higher order measures of robustness can be obtained as we show this in this section. Consider a general quantum map $\mathcal{E} = \mathcal{E}_0 + \eta \mathcal{E}_1$ with $\eta \ll 1$. We aim at finding its spectrum defined via the equations:
\begin{align}
\bbra{\text{l}}\mathcal{E}  = \bbra{\text{l}} \varepsilon, \quad \& \quad \mathcal{E} \kket{\text{r}} = \varepsilon \kket{\text{r}},\label{eq:all_orders}
\end{align}
for a particular eigenvalue $\varepsilon$. We assume that $\text{r}$, $\text{l}$ and $\varepsilon$ can be expanded in powers of $\eta$
\begin{align}
\kket{\text{r}} = \sum_{k\geq 0} \eta^k \kket{\text{r}_k}, \qquad 
\bbra{\text{l}} = \sum_{k\geq 0} \eta^k \bbra{\text{l}_k}, \qquad 
\varepsilon = \sum_{k\geq 0} \eta^k \varepsilon_k.
 \end{align}
It follows that Eq.~\eqref{eq:all_orders} can be written
\begin{align}
\mathcal{E}_0 \kket{\text{r}_0} + \sum_{k\geq 1} \eta^{k} \left(\mathcal{E}_0\kket{\text{r}_k} +\mathcal{E}_1\kket{\text{r}_{k-1}} \right) = \sum_{k\geq 1}\sum_{l = 1}^{k} \eta^{k} \varepsilon_l \kket{\text{r}_{k-l}} + \sum_{k\geq 0} \eta_k \varepsilon_0 \kket{\text{r}_k},
\end{align}
which leads to the recurrence relation:
\begin{align}
\left(\mathcal{E}_0 -\lambda_0 \right)\kket{\text{r}_k} = \mathcal{E}_1\kket{\text{r}_{k-1}} -\sum_{l=1}^k \varepsilon_l\kket{\text{r}_{k-l}}.\label{eq:recurrence}
\end{align}
The correction to the eigenvalues can be computed by projecting onto $\bbra{\text{l}_0}$, 
\begin{align}
\varepsilon_k = \qqav{\text{l}_0|\mathcal{E}_1|\text{r}_{k-1}} -\sum_{l=1}^{k-1} \varepsilon_{l}\qqav{\text{l}_0|\text{r}_{k-l}},
\end{align}
with the relation
\begin{align}
\varepsilon_k = \frac{1}{k!}\left|\frac{\partial^k \varepsilon}{\partial^k \eta}\right|_{\eta=0}.
\end{align}
Perturbed eigenvectors can be also computed from Eq.~\eqref{eq:recurrence}, however its expression is quite involved and dependent on the choice of the inverse of the operator $\mathcal{E}_0-\lambda_0$ \cite{Li2014} and we do not include them here. 

\section{Few level systems }\label{app:examples}

Here we include complementary calculations of the few level systems shown in Sec.~\ref{sec:intuition}.\\

\textbf{Single qubit: dephasing}\\
The kicked protocol in Subsec.~\ref{subsec:dephasing} is not robust to adding a perpendicular Hamiltonian since the deformation $\text{H}_S(\eta) = \eta \text{X}/2$ opens a linear gap in the Liouvillian spectrum of Eq.~ \eqref{eq:bistable_qubit}
\begin{align}
\{\lambda_\mu \} &= \acom{0,-\kappa + \sqrt{\kappa^2 -\eta^2}, - \kappa - \sqrt{\kappa^2 -\eta^2}, -2\kappa}\nonumber\\
&\approx \acom{0,-\frac{\eta}{2}, -2 \kappa + \frac{\eta}{2}, -2\kappa} +\mathcal{O}(\eta^2),
\end{align}
and bistability is lost. Since $\chi^{(1)} = 1/2 \not \ll 1$.\\

\textbf{Two qubits: suppression by jump}\\
We now look at the effect of the decay operators within the $\phi$-space. Consider a jump $\text{L}(\eta) = \sum_\alpha \ketbra{\psi_\alpha}{\phi_\alpha} + \eta \, \vec{n}\cdot\vec{\sigma}_\phi$, where $\vec{n}$ is the Bloch vector and $\vec{\sigma}_\phi$ is the Pauli vector in the $\phi$ subspace (e.g. $\vec{\sigma}_\phi^3 = \sum_\alpha (-)^\alpha \ketbra{\phi_\alpha}{\phi_\alpha}$). This gives rise to a suppression factor for the coherences that depends on the particular choice of $\vec{n}$. In general, 
\begin{align}
\text{j}_{\alpha\beta} = \ketbra{\psi_\alpha}{\phi_\beta} + \frac{|\eta|^2}{1+2|\eta|^2} (\vec{n} \cdot \vec{\sigma}_\phi) \ketbra{\phi_\alpha}{\phi_\beta} (\vec{n} \cdot \vec{\sigma}_\phi)  + \frac{1+|\eta|^2}{1+2|\eta|^2} \ketbra{\phi_\alpha}{\phi_\beta}.
\end{align}
For instance, the particular choice of $\vec{n} = (0,0,1)$ leads to
\begin{align}
 \text{j}_{\alpha\beta} = \ketbra{\psi_\alpha}{\psi_\beta} + \frac{1+|\eta|^2(1+(-)^{\alpha+\beta})}{1+2|\eta|^2}\ketbra{\phi_\alpha}{\phi_\beta},
\end{align}
which indicates that dephasing within the decay space is translated into a coherence suppression of order $\sim |\eta|^{-2}$ within the $\phi$ -block.\\

\textbf{Two qubits: suppression by hamiltonian}\\
Finally, we look at the effect of having a Hamiltonian that acts {\color{black} independently} in the $\psi$ and $\phi$ subspaces. This translates into having residual Hamiltonian evolution in the steady-state. Consider, 
\begin{align}
\text{H}_\text{S} = \frac{h}{2} \text{Z}_\psi + \frac{h+\delta}{2} \text{Z}_\phi,
\end{align} 
and $\text{L} = \sum_\alpha \ketbra{\psi_\alpha}{\phi_\alpha}$. The conserved quantities now read
\begin{align}
 \text{j}_{\alpha\beta} = \ketbra{\psi_\alpha}{\psi_\beta} + \frac{\ketbra{\phi_\alpha}{\phi_\beta}}{1-(-)^\alpha(1-\delta_{\alpha\beta}) i\delta }.
\end{align}
Hence, the effect of a coherent evolution within the asymptotic and decay space is to suppress the coherences between the $\phi$-block. However, the strength depends on the effective detuning $\delta$ between the two spaces. 

\section{XY chain: Diagonalization procedure}\label{app:diagonalization}

First we note that $\text{H}_\xi$ can be broken in parity sectors since $[\text{P},\text{H}_\xi] = 0$ and $\text{P} = \prod_r \text{Z}_r$ with possible eigenvalues $p = \pm 1$. With the Jordan-Wigner transformation, we map spins into fermions using:
\begin{align}
\text{Z}_s \leftrightarrow 1-2 \text{c}^\dagger_s \text{c}_s \quad\& \quad \sigma^+_s \leftrightarrow \exp\left(i \pi \sum_{r < s} \text{c}_r^\dagger \text{c}_r\right) \text{c}_s.
\end{align}
Note that this transformation does not depend on $\xi$. After some manipulation and imposing the appropriate boundary conditions in each parity sector we find: 
\begin{align}
\text{H}^\pm_\xi = -J\sum_{r=1}^{L} \left(\text{c}_{r}^\dagger \text{c}_{r+1} + \gamma \text{c}_{r}^\dagger\text{c}_{r+1}^\dagger - h \text{c}_{r}^\dagger\text{c}_{r} + \text{h.c.} \right) - J h L. \label{eq:jwt_ham}
\end{align}
We now take advantage of the fact that $\text{H}^\pm_\xi$ are translationally invariant and perform the Fourier transform
\begin{align}
\text{c}_r = \frac{e^{i\pi /4}}{\sqrt{L}} \sum_{q \in \text{BZ}_{\pm}} e^{i \frac{2\pi}{L} q r} \text{c}_{q}  \quad \& \quad \text{c}_q = \frac{e^{-i\pi /4}}{\sqrt{L}} \sum_{r = 1}^{L} e^{-i \frac{2\pi}{L} q r} \text{c}_{r}.
\end{align}
where $\text{BZ}_{\pm}$ stands for the Brillouin zone specified by:
\begin{align}
& \text{BZ}_+ = \{q = m +\frac{1}{2}\quad m \in \{-L/2, \cdots, L/2-1 \} \} & \text{for } \text{H}_\xi^+, \nonumber \\ 
& \text{BZ}_- = \{q =  \pm m  \quad m \in \{-L/2, \cdots, L/2-1 \} \} \,  & \text{for } \text{H}_\xi^- .
\end{align}
This brings the Hamiltonian to the explicit block quadratic form
\begin{align}
\text{H}_\xi^{\pm} &= J \sum_{q \in \text{BZ}_\pm}  
\begin{pmatrix}
\text{c}_q^\dagger & \text{c}_{-q} 
\end{pmatrix}
\begin{pmatrix}
h-\cos (\frac{2\pi}{L}q) & -\gamma \sin (\frac{2\pi}{L}q)\\
-\gamma \sin(\frac{2\pi}{L}q) & \cos (\frac{2\pi}{L}q)-h
\end{pmatrix}
\begin{pmatrix}
\text{c}_q \\ \text{c}_{-q} ^\dagger
\end{pmatrix},\nonumber\\
& \coloneqq \sum_{q \in \text{BZ}_\pm}  \begin{pmatrix}
\text{c}_q^\dagger & \text{c}_{-q} 
\end{pmatrix}
\text{H}_{\xi,q}
\begin{pmatrix}
\text{c}_q \\ \text{c}_{-q} ^\dagger
\end{pmatrix}. \label{eq:matrixham}
\end{align}
We can now perform the $\xi$-dependent Bogoliubov transformation to diagonalize $\text{H}_\xi^{\pm}$. Since the matrix $\text{H}_{\xi,q}$ is a combination of the Pauli matrices in the x and z directions, it can be diagonalized via a rotation
\begin{align}
\text{R}_{\xi,q} = \exp\left(-i \frac{\theta_{\xi,q} }{2} \sigma_{y}\right) = \begin{pmatrix}
\cos\frac{\theta_{\xi,q}}{2} & -\sin\frac{\theta_{\xi,q}}{2} \\
\sin\frac{\theta_{\xi,q}}{2} & \cos\frac{\theta_{\xi,q}}{2} \\
\end{pmatrix},
\end{align}
such that $\tilde{\text{H}}_{\xi,q} = \text{R}_{\xi,q} \text{H}_{\xi,q} \text{R}_{\xi,q} ^\dagger$ is diagonal. Setting the off-diagonal terms to zero requires:
\begin{align}
&\theta_{\xi,q} = \tan^{-1} \frac{\gamma \sin\left(\frac{2\pi}{N} q\right)}{h-\cos\left(\frac{2\pi}{N} q\right)}.
\end{align}
Finally, the Hamiltonian takes the expression:
\begin{align}
\text{H}_\xi^{\pm} &= \frac{1}{2} \sum_{q \in \text{BZ}_\pm} E_{\xi,q} \left(\text{d}_{\xi, q}^\dagger \text{d}_{\xi, q}-\text{d}_{\xi, -q} \text{d}_{\xi, -q}^\dagger \right) = \sum_{q\in \text{BZ}_\pm} E_{\xi,q} (\text{d}_{\xi, q}^\dagger \text{d}_{\xi, q} -1/2),\label{eq:diagonal_ham}
\end{align}
where $\pm$ stands for the even and odd parity sectors, $\text{d}_{\xi,q}$ are the Bogoulibov fermions and the dispersion is given by:
\begin{align}
&E_{\xi,q} =E_{\xi,-q} = 2 J \sqrt{\left(h-\cos\left(\frac{2\pi}{L}q\right)\right)^2+\left(\gamma \sin\left(\frac{2\pi}{L}q\right)\right)^2}. \label{eq:spectrum}
\end{align}
Since $\omega_{\xi,q} >0$, the ground state of $\text{H}_\xi^{\pm}$ corresponds to the vacuum of Bogoulibov fermions in each parity sector with a small difference. The ground state of the even parity block corresponds to $\ket{+,\text{GS}} = \bigotimes_{q\in\text{BZ}_+}\ket{0}_q$, where $\text{d}_{\xi,q}\ket{0}_q =0$. In the odd parity sector, there is always an odd number of fermions, and therefore the ground state is given by $\ket{-,\text{GS}} = \text{d}^\dagger_{\xi,0} \bigotimes_{q\in\text{BZ}_-}\ket{0}_q $ (see \cite{Franchini2017}).

\section{XY chain: A sub-manyfold of product ground states}\label{app:product_gs}

There exists a particular sub-manyfold $\xi_p$ of the parameter space $\xi$ (within the ordered phase $h<1$) such that the ground space of the system can be analytically found as product of rotated spin states. This manyfold is often refered as the factorization line. This sub-manyfold is described by the radius-one circle $h^2 + \gamma^2 = 1$, or equivalently, $\xi_p = (J, \sqrt{1-h^2},h) $\cite{Franchini2017}. The exactly degenerated ground states are found: 
\begin{align}
\ket{k,\text{GS}} = \prod_{r=1}^{L}\left(\cos\zeta \ket{\uparrow}_r +(-)^k \sin\zeta\ket{\downarrow}_r\right) \quad \text{for } k=0,1, \label{eq:prod_submanyfold}
\end{align}
where $\cos^2(2\zeta) = (1-\gamma)/(1+\gamma)$. Note that this two states are linearly independent but not orthogonal. Also, both states are connected via $\text{P} \ket{k, \text{GS}} = \ket{k\oplus 1, \text{GS}}$ where $\oplus$ means sum modulo two. This allows to define a ground state within each parity sector as
\begin{align}
\ket{p, \text{GS}}\propto \ket{0,\text{GS}} + p\ket{1,\text{GS}}, 
\end{align}
such that $\text{P}\ket{p, \text{GS}} = p \ket{p,\text{GS}}$. The associated dispersion relation is for each of these ground states is:
\begin{align}
E_{\xi_p,q} = J\left| 1- h \cos\left(\frac{2\pi}{L} q \right)\right|.
\end{align}

\section{XY chain: Pseudo-spin representation}\label{app:pseudospin}

We start be rewriting the Hamiltonian of the system summed over only the positive quasi-momentum part. From Eq.~\eqref{eq:diagonal_ham}, we find
\begin{align}
&\text{H}^\pm_\xi = \sum_{q>0} E_{\xi,q} \left(\text{d}_{\xi,q}^\dagger \text{d}_{\xi,q}-\text{d}_{\xi,-q}^\dagger \text{d}_{\xi,-q}\right) + \nonumber\\
&+\frac{(1\mp 1)}{2}\left(E_{\xi,0} (\text{d}_0 \text{d}^\dagger_0 - \text{d}_0 \text{d}^\dagger_0)+ E_{\xi,\pi} (\text{d}_\pi \text{d}^\dagger_\pi - \text{d}_\pi \text{d}^\dagger_\pi)\right)
\end{align}
where $q>0 = \{ 1/2, \cdots, (L-1)/2 \}$, a total of $L/2$ values; and $q>0 = \{ 1, \cdots, L/2-1\}$, a total of $L/2-1$ values, for the even and odd parity sectors respectively. Folding the BZ into the $q>0$ part, the full Fock space for a given quasimomentum $q$ is spanned by the four states $\ket{0}_q$, $\text{c}_q^\dagger \ket{0}_q$,  $\text{c}_{-q}^\dagger \ket{0}_q$, and $\text{c}_q^\dagger \text{c}_{-q}^\dagger \ket{0}_q$. Within the vanishing \textit{total} quasi-momentum subspace, defined by $\text{Q} = \sum_q q \text{d}_q^\dagger \text{d}_q$, the states can be labeled with binary numbers collected in the vector $\vec{s}$, such that
\begin{align}
\ket{\pm,\vec{s}} &= \left\{\begin{array}{l l}
				\prod_{q > 0} \left( \text{c}_q^\dagger \text{c}_{-q}^\dagger\right)^{m_k} \ket{+,\text{vac}} \\
				\text{c}_0^\dagger \prod_{q > 0} \left( \text{c}_q^\dagger \text{c}_{-q}^\dagger\right)^{m_k} \ket{\text{-,vac}},\\
				\end{array}\right. \label{eq:fock_space}
\end{align}
with the subtlety that the $q = -\pi$ mode should be unoccupied for the odd parity sector when $L$ is even. This comes from the fact that the mode $-q = q = -\pi$ goes into itself at the borders of the Brillouin zone. If the system is prepared in a state of the subspace of vanishing \textit{total} quasi-momentum $q$, for instance $\ket{\pm,\text{GS}}$, a pseudo-spin representation is possible for each block $q$. This is because $\text{H}_\xi^\pm$ only connects the Fock vacuum (of the physical fermions) $\ket{\text{vac}}$ with the state $\text{c}_q^\dagger \text{c}_{-q}^\dagger\ket{\text{vac}} $ for each $q$. We introduce the notation:
\begin{align}
\text{c}_q^\dagger \text{c}_{-q}^\dagger \ket{\text{vac}}_q \leftrightarrow \ket{\uparrow}_q \quad \& \quad \ket{\text{vac}}_q \leftrightarrow \ket{\downarrow}_q.
\end{align}
Then for any operator $\text{O} = \sum_q \text{O}_q$, that acts independently on the different subspaces of quasi-momentum $q$, we can decompose it in this basis as:
\begin{align}
 \text{O}_q = \sum_{ss' =\uparrow \downarrow } \qav{s|\text{O}_q|s'} \ketbra{s}{s'}.
\end{align}
In the second-quantization, the expression of $\text{O}$ is given by:
\begin{align}
\text{O} = \sum_{q>0}  
\begin{pmatrix}
\text{c}_q^\dagger & \text{c}_{-q} 
\end{pmatrix}
\begin{pmatrix}
O_{q,\uparrow \uparrow} & O_{q,\uparrow \downarrow}\\
O_{q,\downarrow \uparrow} & O_{q,\downarrow \downarrow}
\end{pmatrix}
\begin{pmatrix}
\text{c}_q \\ \text{c}_{-q} ^\dagger
\end{pmatrix}.
\end{align}

\section{Detailed derivation of the Master Equation}\label{app:me_derivation}

Our starting point is the system-bath Hamiltonian in Eq.~\eqref{eq:system-bath}. For simplicity of the calculation, we also restrict ourselves to the zero quasi-momentum subspace. For further use, we introduce the interaction picture as $\text{O}(t) = \exp(i \text{H}_\text{S}t) \text{O} \exp(-i \text{H}_\text{S}t)$. Our starting point is the Redfield equation in the interaction picture:
\begin{align}
\dot{\tilde{\rho}}(t) = -\text{tr}_B \int_0^\infty ds \com {\text{H}_B (t), \com{ \text{H}_{SB}(t-s),\tilde{\rho}(t) \otimes \rho_\text{eq}}}, \label{eq:second_order}
\end{align}
where $\rho_{eq}\propto\exp(-\beta \text{H}_B)$. Next, we decompose the magnetization operator according to:
\begin{align}
\text{M}_z &= \sum_{r}(1-2\text{c}^\dagger_r\text{c}_r) = \sum_q  (1-2\text{c}^\dagger_q\text{c}_q) =  \sum_{q} (\text{c}^\dagger_q\text{c}_q - \text{c}_{-q}\text{c}^\dagger_{-q}) \nonumber\\ 
&=  \sum_{q}  
\begin{pmatrix}
\text{d}_q^\dagger & \text{d}_{-q} 
\end{pmatrix}
\begin{pmatrix}
\cos (\theta_{\xi,q}) & \sin (\theta_{\xi,q})\\
 \sin (\theta_{\xi,q}) & -\cos (\theta_{\xi,q})
\end{pmatrix}
\begin{pmatrix}
\text{d}_q \\ \text{d}_{-q} ^\dagger
\end{pmatrix} .
\end{align}
Note that this gives rise to the expression introduced in Eq.~\eqref{eq:mz_partition} of the main text in terms of the jump operators $\text{L}_{\xi,q}^\alpha$. Using this expression in Eq.~\eqref{eq:second_order} it follows
\begin{align}
\dot{\tilde{\rho}}(t) = \epsilon^2\sum_{q, q'} \sum_{\alpha,\alpha'} \int_0^\infty ds \left(\text{L}_{\xi,q'}^{\alpha}(t-s)\tilde{\rho}(t)\text{L}^{\alpha\dagger}_{\xi,q}(t) - \text{L}_{\xi,q}^{\alpha\dagger}(t) \text{L}_{\xi,q'}^{\alpha}(t-s) \tilde{\rho}(t)\right)C(s) + \text{h.c.},
\end{align}
where $C(t-t') = \text{tr}_\text{B}[\text{B}(t)\text{B}(t')\rho_\text{eq}]$. Note that, by construction $\text{L}_{\xi,q}^\alpha(t) = \text{L}_{\xi,q}^\alpha(0)\exp(i \omega_{\xi,q}^\alpha t)$ and $\omega_{\xi,q}^0 = 0$, $\omega_{\xi,q}^{\uparrow \downarrow}= \pm 2E_{\xi,q}$ and then
\begin{align}
\dot{\tilde{\rho}}(t) = -\epsilon^2\sum_{q, q'}\sum_{\alpha,\alpha'} \Gamma(\omega_{\xi,q}^\alpha) e^{-i(\omega_{\xi,q}^\alpha - \omega_{\xi,q'}^{\alpha'})t } \left( \text{L}^{\alpha\dagger}_{\xi,q} \text{L}^{\alpha'}_{\xi,q'}\tilde{\rho}(t) -  \text{L}^{\alpha'}_{\xi,q'}\tilde{\rho}(t)\text{L}^{\alpha\dagger}_{\xi,q}\right) + \text{h.c.}. \label{eq:born_markov}
\end{align}
where we have introduced $\Gamma(\omega) = \int_0^\infty ds\, C(s) \exp(i \omega s)$. We take the real and imaginary parts of $\Gamma(\omega)$ as follows:
\begin{align}
\Gamma(\omega) &= \frac{1}{2}\kappa(\omega) + i S(\omega), \nonumber \\
\kappa(\omega) &= \int_0^\infty ds C(s) e^{i\omega s} +\int_0^\infty C^*(s) e^{-i\omega s} = \int_{-\infty}^\infty ds\, C(s) e^{i\omega s} = C(\omega), \nonumber \\
S(\omega) &= \frac{1}{2i} \left(\int_0^\infty ds C(s) e^{i\omega s} -\int_0^\infty ds C^*(s) e^{-i\omega s}\right).
\end{align}
Then, we can compute the bath correlation function as: 
\begin{align}
C(s) &= \sum_{k k',\mu \mu'}\qav{\left( g_{k,\mu} \text{b}^\dagger_{k,\mu} e^{i\Omega_{k,\mu} s} + g_{k,\mu}^* \text{b}_{k,\mu} e^{-i\Omega_{\alpha,\mu} s}  \right) \left(g_{k',\mu'} \text{b}^\dagger_{k',\mu'} + g_{k',\mu'}^* \text{b}_{\alpha',\mu'}\right)  \rho_\text{eq}} \nonumber \\
&= \sum_{k,\mu} |g_{k,\mu}|^2\left( n_{k}(\Omega_{k,\mu}) e^{i\Omega_{k,\mu} s} + (1 + n_{k}(\Omega_{k,\mu}) )e^{-i\Omega_{k,\mu} s}\right) \nonumber\\
&= \sum_k \int_0^\infty \frac{d\omega}{2\pi} J_{k}(\omega)  \left( n_{k}(\omega) e^{i\omega s} + (1 + n_{k}(\omega) )e^{-i\omega s}\right),
\end{align}
where we have introduced the spectral density $J(\omega) = 2\pi \sum_{k} |g_{k}|^2 \delta(\omega-\Omega_{k})$ {\color{black} and the bosonic occupation number $n(\omega) = (\exp(\beta \omega)-1)^{-1}$}.  When the number of the modes of the bath tends to infinity, the spectral function is usually approximated with a continuous continuous function of $\omega$.  The spectral function $J(\omega)$ can be analytically continued towards negative frequencies as $J(-\omega) = -J(\omega)$, which allows:
\begin{align}
C(s) = \int_{-\infty}^\infty \frac{d\omega}{2\pi} J(\omega) (1+n(\omega) ) e^{-i\omega s},
\end{align}
and reading of $C(\omega) = J(\omega) (1  + n_k(\omega) )$. The function $S(\omega)$ is given
\begin{align}
S(\omega) = \text{Im}\com{ \int_\mathbb{R} \frac{d\omega'}{2\pi} J (\omega') (n(\omega') +1) \int_0^\infty ds e^{i (\omega-\omega')s} }= \mathcal{P} \int_\mathbb{R} \frac{d\omega'}{2\pi} \frac{J (\omega') (n(\omega') +1)}{\omega-\omega'},
\end{align}
and it is responsible for the Lamb shift. From now on, we ignore it which corresponds to substituting $\Gamma(\omega) \to 1/2 \kappa(\omega)$ in Eq.~\eqref{eq:born_markov}. Note that, since $\kappa(\omega=0)=0$, the jump operators $\text{L}_{\xi,q}^0$ do not contribute to dissipation at zero temperature. For finite temperature, they do not contribute either for super-ohmic spectral densities, which we assume to be our case. 
\begin{align}
\dot{\tilde{\rho}}(t) = -\frac{\epsilon^2}{2}\sum_{q, q'}\sum_{\alpha,\alpha' \neq 0} \kappa(\omega_{\xi,q}^\alpha) e^{-i(\omega_{\xi,q}^\alpha - \omega_{\xi,q'}^{\alpha'})t } \left( \text{L}^{\alpha\dagger}_{\xi,q} \text{L}^{\alpha'}_{\xi,q'}\tilde{\rho}(t) -  \text{L}^{\alpha'}_{\xi,q'}\tilde{\rho}(t)\text{L}^{\alpha\dagger}_{\xi,q}\right) + \text{h.c.}
\end{align}
The next step consists in using the secular approximation, that selects only those terms that fulfill the resonant condition $\omega^\alpha_{\xi,q} - \omega^{\alpha'}_{\xi,q'} = 0$. As explained in Sec.~\ref{sec:openXY}, two different situations can arise leading to the equations Eq.~\eqref{eq:liouv_xy_coll} and Eq.~\eqref{eq:liouv_xy_loc} of the main text.

%

\section{Numerical ME}\label{app:num_me}

{\color{black} The ME derived in Sec.~\eqref{sec:openXY} gives intuition about the open system dynamics of the XY chain but is hard to implement numerically. Here, we take a different approach leading to a more numerically tractable equation. We start from a total system-bath Hamiltonian in the form $\text{H} = \text{H}_\xi + \text{H}_\text{SB} + \text{H}_\text{B}$ where $\text{H}_\text{SB} = \text{M}_z \otimes \text{B}$ and $\text{M}_z = \sum_r \text{Z}_r$. This is brought into 
\begin{align}
\dot\rho = -i [\text{H}_\xi,\rho] + \int_0^\infty ds \text{tr}_\text{E} \left(\text{H}_\text{SB} \rho\otimes\rho_\text{eq.} \text{H}_\text{SB}(-s) - \rho \otimes \rho_\text{eq.}\text{H}_\text{SB}(-s) \text{H}_\text{SB}+\text{h.c.} \right),
\end{align}
from where we find 
\begin{align}
\dot\rho = -i [\text{H}_\xi,\rho] + \int_0^\infty ds \qav{\text{B}(-s) \text{B}} \left( \text{M}_z \rho \text{M}_z(-s) - \rho \text{M}_z(-s) \text{M}_z \right) +\text{h.c.}
\end{align}
We can now compute the correlation function
\begin{align}
\qav{\text{B}(-s) \text{B}} &= \sum_{kk'} \qav{(g_k \text{b}_k^\dagger e^{-i \omega_k s} +g_k^*\text{b}_k e^{i \omega_k s} \right)\left(g_{k'} \text{b}^\dagger_{k'}+g_{k'}^*\text{b}_{k'})}\nonumber\\
&=\sum_{k} |g_k|^2 \left( \qav{\text{b}^\dagger_k \text{b}_k} e^{-i \omega_k s} +  \qav{\text{b}_k \text{b}^\dagger_k} e^{i\omega_k s}\right) \nonumber\\
&= \int_{\mathbb{R}} \frac{d\omega}{2\pi} J(\omega) (n(\omega) + 1)e^{i\omega s}.
\end{align}
We introduce the complete basis $\{ \ket{k} \}$ for the system Hamiltonian, i.e. $\text{H}_\xi\ket{k} = E_{\xi,k} \ket{k}$. Also we define $\omega_{kl} = E_{\xi, k}-E_{\xi, l}$  and decompose $\text{M}_z(-s) = \sum_{kl} m_{kl} e^{i \omega_{kl}(-s)} \ket{k}\bra{l}$. Then, 
\begin{align}
\dot\rho &= -i [\text{H}_\xi,\rho] + \nonumber\\
&\sum_{kl} \int_{\mathbb{R}} \frac{d\omega}{2\pi} \int_0^\infty ds  J(\omega) (n(\omega) + 1)e^{i(\omega -\omega_{kl})s} \left( \text{M}_z \rho m_{kl}\ket{k}\bra{l} -\rho m_{kl}\ket{k}\bra{l} \text{M}_z \right) +\text{h.c.}
\end{align}
The integration over time can be done using Sokhotski–Plemelj theorem which stands that $\int_{0}^\infty ds \exp(i x s) = \pi \delta(x) -i \mathcal{P}(1/x)$. As often we disregard the principal value term and get the simple equation
\begin{equation}
\dot\rho = \mathcal{L}(\rho) =  -i [\text{H}_\xi,\rho]+ \left[\text{M}_z, \rho \text{D} \right] + \left[ \text{D}^\dagger \rho, \text{M}_z\right],
\end{equation}
where the operator $\text{D} =1/2 \sum_{kl} J(\omega_{kl}) (n(\omega_{kl}) + 1) m_{kl} \ket{k}\bra{l} $.}

\bibliographystyle{unsrtnat}
\bibliography{time_crystals_doi}

\end{document}